\documentclass[aps,prx,twocolumn,longbibliography,eqsecnum,superscriptaddress]{revtex4-2}

\usepackage{amsmath,amsfonts, amssymb}
\usepackage{bm}
\usepackage{graphicx}
\usepackage{verbatim}
\usepackage{hyperref}
\usepackage{xcolor}
\usepackage{enumerate}
\usepackage{tikz,cleveref}
\usepackage{bbold}
\usepackage{booktabs}
\usepackage{multirow}
\usepackage{multibib}
\usepackage{sidecap}

\usepackage[T1]{fontenc}
\usepackage[utf8]{inputenc}
\usepackage{newtxtext}
\usepackage{amsmath}
\usepackage{mathtools}
\usepackage{amsfonts}
\usepackage{bm}
\usepackage{amssymb}

\usepackage{newtxmath}

\usepackage[utf8]{inputenc}
\usepackage[T1]{fontenc}
\usepackage{xcolor}

\usepackage{tabularx}

\usepackage{bm}

\pdfoutput=1
\usepackage{color}
\definecolor{LinkColor}{rgb}{0.156,0.439,0.688}
\usepackage{hyperref}
\hypersetup{
colorlinks=true,
citecolor=LinkColor,
linkcolor=LinkColor,
urlcolor=LinkColor
}
\usepackage{multirow}

\definecolor{todocolor}{rgb}{0.76, 0.0, 0.03}

\newcommand{\avg}{\mathrm{avg}}
\newcommand{\typ}{\mathrm{typ}}

\newcommand{\C}{\bm{\mathcal{{C}}}}
\usepackage{tikz}
\usepackage{textcomp}

\def\be{\begin{equation}}
\def\ee{\end{equation}}
\def\bea{\begin{eqnarray}}
\def\eea{\end{eqnarray}}

\begin{document}
\title{Statistics of systemwide correlations in the random-field XXZ chain:\\
Importance of rare events in the many-body localized phase}
\author{Jeanne Colbois}
\email{physics.jeannecolbois@gmail.com}
\affiliation{Laboratoire de Physique Th\'{e}orique, Universit\'{e} de Toulouse, CNRS, UPS, France}
\affiliation{MajuLab, CNRS-UCA-SU-NUS-NTU International Joint Research Unit, Singapore}
\affiliation{Department of Materials Science and Engineering and Centre for Quantum Technologies, National University of Singapore, Singapore}
\author{Fabien Alet}
\email{fabien.alet@cnrs.fr}
\affiliation{Laboratoire de Physique Th\'{e}orique, Universit\'{e} de Toulouse, CNRS, UPS, France}
\author{Nicolas Laflorencie}
\email{nicolas.laflorencie@cnrs.fr}
\affiliation{Laboratoire de Physique Th\'{e}orique, Universit\'{e} de Toulouse, CNRS, UPS, France}
\date{\today}
\begin{abstract}
Motivated by recent debates around the many-body localization (MBL) problem, and in particular its stability against systemwide resonances, we investigate long-distance spin-spin correlations across the phase diagram of the random-field XXZ model, with a particular focus on the strong disorder regime. Building on state-of-the-art shift-invert diagonalization techniques, we study the high-energy behavior of transverse and longitudinal correlation functions, computed at the largest possible distance, for a broad range of disorder and interaction strengths.  Our results show that while transverse correlations display a fairly stable exponential decay over the entire XXZ phase diagram, longitudinal correlations exhibit markedly different behavior, revealing distinct physical regimes. More precisely, we identify an intermediate disorder region where standard observables show well-converged MBL behavior [J. Colbois {\it{et al}}., \href{https://journals.aps.org/prl/abstract/10.1103/PhysRevLett.133.116502}{Phys. Rev. Lett. 133, 116502 (2024})] while the distributions of longitudinal correlations reveal unexpected fat-tails towards large values. These rare events strongly influence the average decay of longitudinal correlations, which we find to be algebraic in a broad region inside the supposed MBL phase, whereas the typical decay remains mostly exponential.  At stronger disorder and weaker interactions, this intermediate regime is replaced by a more conventional exponential decay with short correlation lengths for both typical and average correlators, as expected for standard localization. Our findings shed light on the systemwide instabilities and raise important questions about the impact of such rare but large long-range correlations on the stability of the MBL phase. Finally, we discuss the possible fate of the intermediate region in the context of recent perspectives in the field.
\end{abstract}
\maketitle
\tableofcontents
\section{Introduction}
Over the past decades, the many-body localization (MBL) phenomenon has attracted a great deal of interest, leading to a significant effort deployed to understand the interacting version of an Anderson insulator~\cite{fleishman_interactions_1980,altshuler_quasiparticle_1997,jacquod_emergence_1997,gornyi_interacting_2005,basko_metalinsulator_2006,oganesyan_localization_2007,nandkishore_many-body_2015,alet_many-body_2018,abanin_colloquium_2019,sierant_many-body_2024}. 
Among the crucial points that have been discussed  recently, the interpretation of finite-size and/or finite-time numerical simulations plays a central role~\cite{doggen_many-body_2018,weiner_slow_2019,panda_can_2020,chanda_time_2020,sierant_thouless_2020,abanin_distinguishing_2021,doggen_many-body_2021,sierant_challenges_2022,morningstar_avalanches_2022}. 
Indeed, our current understanding of the MBL problem has been strongly shaped by the results of simulations of relatively small interacting quantum disordered systems.

Nevertheless, it is quite widely accepted~\cite{znidaric_many-body_2008,pal_many-body_2010,bardarson_unbounded_2012,luca_ergodicity_2013,luitz_many-body_2015,bera_many-body_2015,bar_lev_absence_2015,serbyn_criterion_2015,gray_many-body_2017,pietracaprina_shift-invert_2018,mace_multifractal_2019,sierant_polynomially_2020,laflorencie_chain_2020,roy_fock-space_2021}  that in one dimension, strong enough disorder favors a stable MBL phase or regime, on the length scales (a few tens of spins) probed by these simulations. However, after almost two decades of very hard work, there is still no clear consensus on some aspects, not least being the very existence of the MBL phase in the thermodynamic limit~\cite{suntajs_quantum_2020,sels_dynamical_2021,sels_thermalization_2023,weisse_operator_2024} , even for the few models that have been intensively studied, such as the random-field Heisenberg chain~\cite{pal_many-body_2010,luitz_many-body_2015,morningstar_avalanches_2022}. One should also mention the few attempts to mathematically prove the existence of the MBL as a new state of matter preventing thermal equilibrium~\cite{imbrie_many-body_2016,imbrie_diagonalization_2016,yin_eigenstate_2024,deroeck_absence_2024}.
Interestingly, they have all focused on a certain class of models of quantum spin $1/2$ chains with no symmetry, in contrast to the by-now standard U(1) symmetric random-field Heisenberg chain. In addition to these theoretical results, experimental evidences of localization have been reported by several groups~\cite{schreiber_observation_2015,smith_many-body_2016,choi_exploring_2016,luschen_observation_2017,roushan_spectroscopic_2017,lukin_probing_2019,rispoli_quantum_2019,guo_observation_2021,leonard_probing_2023}, reinforcing the idea that the MBL phase can exist, but also raising the same questions about finite size and finite time limitations. 
\subsection{Instabilities}
One important aspect in understanding MBL stability is to identify the relevant mechanisms that could destabilize the interacting insulator. This knowledge would then allow a more complete theoretical modeling of the elusive MBL phenomenon, a problem that has recently benefited from important advances such as the theory of avalanches~\cite{de_roeck_stability_2017,thiery_many-body_2018,crowley_avalanche_2020} and of many-body resonances~\cite{villalonga_eigenstates_2020,garratt_local_2021,garratt_resonant_2022,crowley_constructive_2022,Ha_many-body_2023,long_phenomenology_2023}. 
Attempts at observing or modeling thermalization through avalanches, as triggered by an ergodic inclusion in an otherwise strongly localized background, have taken several forms. Idealized setups modeling the ergodic inclusion as a random matrix~\cite{luitz_how_2017,suntajs_ergodicity_2022,pawlik_many-body_2024,suntajs_similarity_2024} are consistent with theoretical predictions~\cite{de_roeck_stability_2017,thiery_many-body_2018}, even for small systems. Other setups~\cite{sels_bath-induced_2022,morningstar_avalanches_2022}, assuming the avalanche has started and modeling its action through a weakly-coupled bath, find a critical disorder much larger than those previously obtained using standard observables~\cite{luitz_many-body_2015,pietracaprina_shift-invert_2018,mace_multifractal_2019,sierant_polynomially_2020, laflorencie_chain_2020,colbois_interaction_2024}, while with a similar setup Ref.~\cite{scocco_2024_thermalization} reports a larger range of stability of MBL. Investigations of more realistic models of local thermal inclusions~\cite{peacock_many-body_2023,colmenarez_ergodic_2024,szoldra_catching_2024} reach different conclusions, some finding no or little evidence of effective thermalization by the inclusion~\cite{colmenarez_ergodic_2024,szoldra_catching_2024}, some concluding in favor of thermalization albeit in a different model~\cite{peacock_many-body_2023}. Remarkably, these ideas have been tested experimentally~\cite{leonard_probing_2023}, but again on small chains, making the interpretation in terms of avalanches quite delicate. 

Another type of MBL instabilities actively discussed are many-body resonances which, loosely speaking, correspond to the hybridization by interaction terms of ``localized'' many-body eigenstates differing by spin flips located in a certain spatial range in real-space~\cite{gopalakrishnan_low_2015,imbrie_diagonalization_2016, garratt_local_2021,villalonga_eigenstates_2020,crowley_constructive_2022}. The ``localized'' many-body eigenstates that participate the resonance can be adiabatically connected to the large disorder limit (where a local integral of motions picture apply~\cite{huse_phenomenology_2014,serbyn_local_2013,ros_integrals_2015}), and have an energy difference exponentially small with the resonance range~\cite{crowley_constructive_2022}. Long-range resonances imply hybridization between states that differ by an extensive amount (a fraction of the system size), and are extremely close in energy. Such many-body resonances are at the heart of the formal proofs of stability of MBL~\cite{imbrie_diagonalization_2016,imbrie_many-body_2016,deroeck_absence_2024} and are thought to destabilize the MBL phase. Several prior works found indirect signs of (long-range) many-body resonances in microscopic MBL models~\cite{gopalakrishnan_low_2015,geraedts_many-body_2016,khemani_critical_2017,colmenarez_statistics_2019,villalonga_characterizing_2020} while more direct approaches~\cite{kjall_many-body_2018,villalonga_eigenstates_2020,garratt_local_2021,garratt_resonant_2022,morningstar_avalanches_2022,Ha_many-body_2023} have also been designed to identify resonances -- most of these coming with a certain level of modeling~\cite{crowley_constructive_2022,long_phenomenology_2023}.

We finally note that these two destabilization mechanisms need not be unrelated, and Ref.~\cite{Ha_many-body_2023} discusses that avalanches primarily propagate through strong, rare, (near-)resonances. The statistics of long-range resonances, including rare events, certainly play a crucial role when approaching the ergodic transition, as been argued {\it e.g.} in Refs.~\cite{monthus_many-body-localization_2016,biroli_large-deviation_2024} borrowing ideas from glassy physics. Nevertheless, a good understanding of systemwide resonances is not yet fully achieved, especially at the microscopic level. Our work aims to present a contribution to this topic through a comprehensive study of simple objects, namely long-range spin correlation functions in eigenstates, which have the potential to capture long-range resonances and allow us to provide a finer insight into the MBL phase diagram, as we show below.
\subsection{Spin correlations}

 In the vast literature on MBL and its lattice simulations in U(1) symmetric models, only a handful of papers focused on pairwise correlation functions in eigenstates~\cite{pal_many-body_2010,lim_many-body_2016,colmenarez_statistics_2019,villalonga_characterizing_2020, Kulshreshtha_approximating_2019, varma_length_2019,hemery_identifying_2022}, despite the fact that they can be routinely measured in several experimental platforms~\cite{smith_many-body_2016,choi_exploring_2016,lukin_probing_2019}. In practice, the main efforts have focused on the two-site quantum mutual information~\cite{villalonga_characterizing_2020,morningstar_avalanches_2022}, an entanglement estimate that directly depends on single-site and two-body correlations~\cite{tomasi_quantum_2017}.

In the present work, we aim to further explore the phase diagram of a prototypical model hosting MBL with the help of long-distance correlations, extending in several ways our recent work~\cite{colbois_interaction_2024}. The random-field spin-1/2 XXZ model is defined on a periodic chain of length $L$, by the Hamiltonian
\be
    {\cal{H}}_{\Delta}=\sum_{i=1}^{L}\left(S_i^x S_{i+1}^{x} + S_i^y S_{i+1}^{y} + \Delta S_i^z S_{i+1}^{z}+h_iS_i^z\right),
    \label{eq:H}
\ee
where $\Delta \in \left[0,1\right]$ controls the interaction (the Ising anisotropy), and the random fields are drawn from a uniform distribution, $h_i \in [-h,h]$. 
Crucially, ${\cal{H}}_{\Delta}$ preserves the total magnetization $S^z_{\rm{tot}} =\sum_{i=1}^{L} S_i^z$. We highly rely on this U(1) symmetry. 
Throughout this paper, we consider eigenstates of Eq.~\eqref{eq:H} in the middle of the spectrum for the largest sector of fixed total magnetization. As we focus on even sizes $L$, we always consider $S^{z}_{\rm tot}=0$. 
Eigenstates are obtained through shift-invert exact diagonalization (ED)~\cite{pietracaprina_shift-invert_2018} on lattices up to length $L=20$. 
For each value of $\Delta, h$, we consider between 1000 and 5000 realizations $\{ h_i \}$ of disorder.

\begin{SCfigure*}
    \centering
    \includegraphics[width=1.2\columnwidth]{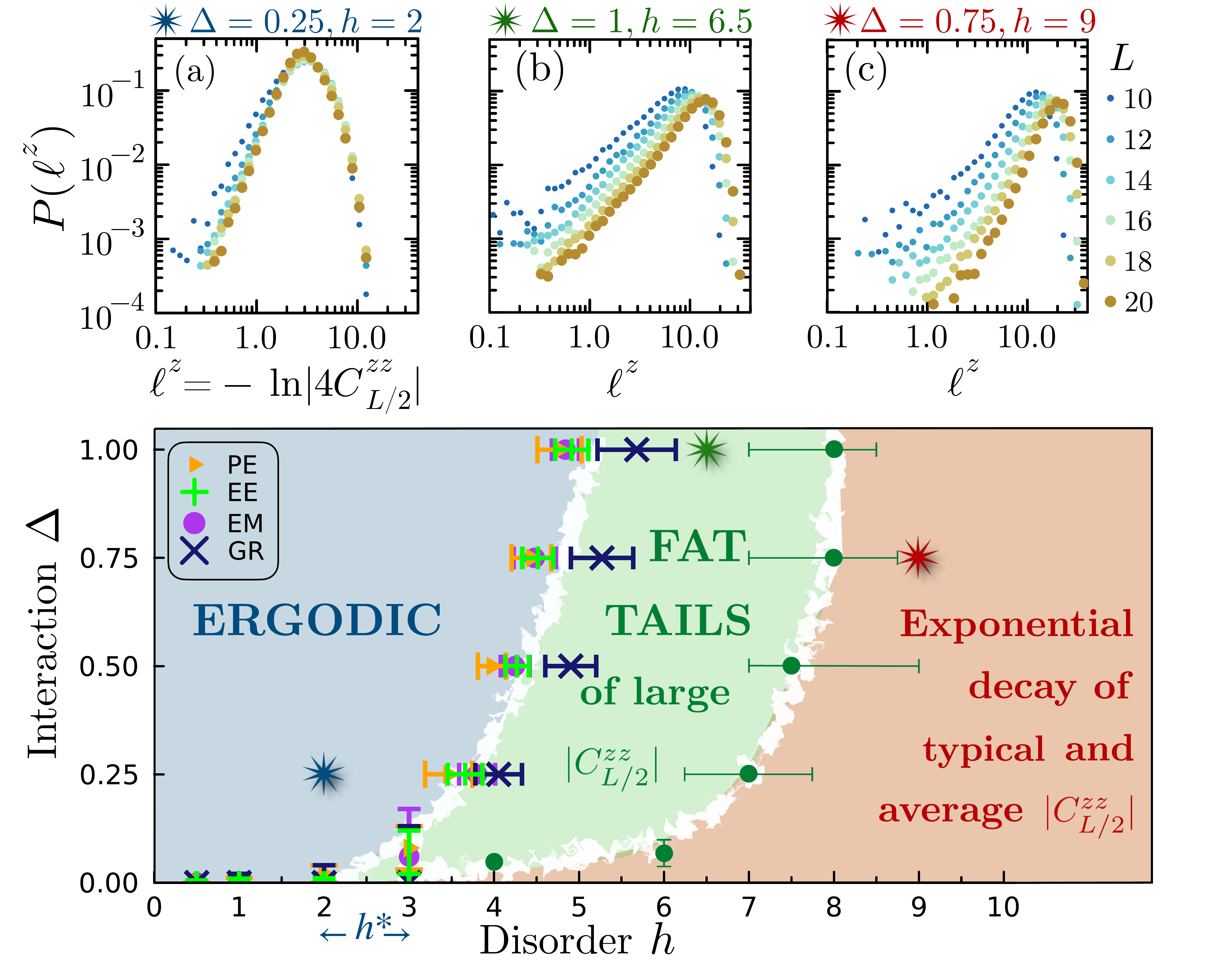}
    \caption{\label{fig:phasediag} High energy phase diagram of the random-field XXZ chain Eq.~\eqref{eq:H} as obtained for mid-spectrum eigenstates from shift-invert diagonalization. The extrapolated transition points out of the ergodic phase (blue region), as previously obtained~\cite{colbois_interaction_2024} using four standard observables, are shown together with new markers (green dots, \emph{not} extrapolated), dividing the phase diagram into three qualitatively different regimes with respect to the midchain longitudinal correlations $C^{zz}_{L/2}$. Between the red and blue regions, our work highlights the importance of a new intermediate regime (green region) where anomalously large correlation events appear in fat-tailed distributions and dominate the average decay of $C^{zz}_{L/2}$. The estimates of the positions of these green markers are discussed in Appendices~\ref{sec:Horizonal} and~\ref{sec:Vertical}. The top insets show the three qualitatively different behaviors for the distributions $P(-\ln|4C_{L/2}^{zz}|)$ for three representative points, indicated by colored stars: (a) self-averaging in the ergodic regime; (b) fat-tail behavior for large correlation events ($\ell^z\to 0$) in the intermediate regime; (c) log-normal-like for strong disorder, where both the average and the typical function decay equally exponentially.}
\end{SCfigure*}

Before presenting the main results of the current paper, let us recall previous findings obtained in Ref.~\cite{colbois_interaction_2024} on the phase diagram for mid-spectrum eigenstates, see Fig.~\ref{fig:phasediag}. 
First, we have compiled evidence showing that the Anderson localized (AL) line is unstable for any finite interaction $\Delta$, leading to an ergodic regime when $h<h^*\sim 2-3$, whereas it connects to the MBL regime for sufficiently large disorder. 
The ergodic to MBL transition line in the disorder $h$ --- interaction $\Delta$ plane, has been obtained through an infinite-size extrapolation ($L\to \infty$, symbols in Fig.~\ref{fig:phasediag}) of finite-size data (up to $L=21$) of four standard observables: participation entropy (PE), entanglement entropy (EE),  extreme magnetizations (EM) and  gap ratio (GR). 
Second, beyond this line, at stronger disorder, we identified a regime of instabilities where, in particular, the spin-spin correlation functions at long-distance ($L/2$) show a non-trivial response upon increasing the system length $L$.

This last point is the main motivation of the present work, where we propose a comprehensive study of the spin-spin correlation functions as a probe across the phase diagram of the random-field XXZ chain model Eq.~\eqref{eq:H} at high energy, shown in Fig.~\ref{fig:phasediag}. 
We denote as follows the connected spin-spin correlation functions
\be
    \label{eq:connected}
    C_{ij}^{\alpha\alpha} = \langle S_i^{\alpha} S_{j}^{\alpha} \rangle  - \langle S_i^{\alpha} \rangle \langle S_{j}^{\alpha} \rangle,
\ee
with $\alpha = x, y, z$ and where the expectation values are calculated for many-body mid-spectrum eigenstates.
The XXZ Hamiltonian Eq.~\eqref{eq:H} commutes with $S^z_{\rm{tot}}$, which implies that the $\alpha = x$ and $\alpha = y$ components, \emph{transverse} with respect to the random field, are equivalent by symmetry, and that $\langle S_i^x\rangle=0$, which is not true for the \emph{longitudinal}  component ($\langle S_i^z\rangle\neq 0$). 

Throughout this work, we mostly use the absolute value of the correlations~\cite{pal_many-body_2010,lim_many-body_2016}, so that we can define the \emph{typical} correlations at distance $r=|i-j|$
\begin{equation}
    \label{eq:typical}
    C_{r,{\rm typ}}^{\alpha \alpha} := \exp(\overline{\ln |C_{r}^{\alpha \alpha} |}),
\end{equation}
where the bar denotes averaging over positions, eigenstates as well as disorder realizations.
Considering the absolute value also in the \emph{average} correlations
\begin{equation}
    \label{eq:average}
    C_{r,{\rm avg}}^{\alpha \alpha} := \overline{|C_{r}^{\alpha \alpha}|}
\end{equation}
facilitates the comparison with the typical behavior, and is helpful for the study of the longitudinal correlations, which are in majority negative in the ergodic regime, and fully negative in the non-interacting limit~\cite{colmenarez_statistics_2019}.
Because of the exponential decay of various correlations throughout the phase diagram, we often display directly their logarithm.

As a probe for long-range resonances, we focus particularly on the \emph{midchain} correlations, between spins at the maximal distance $r=L/2$ in periodic chains. 
We consider their full probability distributions $P(C_{L/2}^{\alpha \alpha})$, and, to emphasize the typical behavior, $P(\ln |C_{L/2}^{\alpha \alpha}|)$. 
The main focus is therefore on the decay of the typical and average midchain correlations for different system sizes $L$, rather than on their decay in the \emph{bulk} as a function of $r$ for a given system size.

\subsection{Main results}

Figure \ref{fig:phasediag} shows the phase diagram of the random-field XXZ chain Eq.~\eqref{eq:H}, across which we have studied in great detail the different behaviors of the spin-spin correlation functions at long distances $C^{\alpha\alpha}_{L/2}$. We thus provide a deep extension of our initial results~\cite{colbois_interaction_2024}, which can be summarized as follows.

\begin{itemize}
    \item[{\it (i)}] The transverse correlation functions $C^{xx}_{L/2}$ show no significant change throughout the phase diagram: they always show an exponential decay, both for their average and for their typical behavior. While the origin of such exponential decays is very different for weak and strong disorder, $C^{xx}_{L/2}$ is not able to distinguish between the different regimes.
    
\item[{\it (ii)}] This is in sharp contrast to the longitudinal correlators $C^{zz}_{L/2}$, which follow distinctly different patterns across the phase diagram: an exponential decay for both typical and average in the strong disorder limit, identified as MBL, precedes a marked difference between typical and average correlators at intermediate disorder, before finally giving rise to the thermal phase characterized by midchain correlations decaying as a power law (due to magnetization conservation) at lower disorder. 

\item[{\it (iii)}]  These different behaviors result from a qualitative change in the probability distribution of the longitudinal correlators. Between the strong and weak disorder limits, we show the existence of a rather broad intermediate regime characterized by a fat tail (presumably power-law) for the large correlations, leading to an average dominated by rare events. In this intermediate region, however, the usual observables (spectral statistics, entanglement, participation, extreme magnetization) show well-converged MBL features~\cite{sierant_polynomially_2020,colbois_interaction_2024}.
\end{itemize}

Our results are strengthened by a detailed study of the two limiting cases of the thermal phase (at low disorder, finite interaction) and the many-body Anderson localized phase (at $\Delta=0$).  In the main text, results along several scans are discussed: the AL line at $\Delta = 0$; the random-field Heisenberg model (RFHM) line at $\Delta = 1$; and  a vertical cut at $h = 6$. Results along other horizontal ($\Delta = 0.25, 0.5, 0.75$) and vertical ($h = 4, 6$) lines are discussed in Appendix.

\subsection{Paper organization}
The rest of the paper is organized as follows. We first discuss in Section~\ref{sec:warmup} the two limit cases: on the one hand, the many-body Anderson insulator ($\Delta=0$), and on the other hand, deep in the ergodic regime, where the eigenstate thermalization hypothesis (ETH)~\cite{deutsch_quantum_1991,srednicki_chaos_1994,rigol_thermalization_2008} applies. 
In particular, we study the joint behavior of the longitudinal and transverse correlation functions in these two regimes, which will serve as a guide for our initial exploration of the phase diagram. 
Section~\ref{sec:Overview} provides an overview of the phase diagram through the lens of midchain correlators following two representative  cuts: one horitzontal along the highly studied random-field Heisenberg model line at $\Delta=1$, and another one along a vertical cut in the phase diagram at $h=6$. 
Typical and average correlations and the related correlation lengths are discussed in detail. Their striking differences motivate a focus on the large correlations, which we discuss in Section~\ref{sec:LargeCorrelations}.
Our results are summarized and put in perspective in Section~\ref{sec:SummaryOutlook} in relation with the recent literature on instabilities and resonances of the MBL phase, along with a brief conclusion.

Our paper is supported by several appendices. Appendix~\ref{sec:fermions} provides further details regarding the free fermions results. Appendices~\ref{sec:RS} and~\ref{sec:variance} discuss respectively the distributions for a random vector in the $S^{z}_{\rm tot}=0$ sector and the variances in the interacting case.
Appendix~\ref{sec:Horizonal} contains a similar analysis as in Sections~\ref{sec:Overview} and~\ref{sec:LargeCorrelations} for horizontal cuts at $\Delta = 0.25$ and $\Delta = 0.5$. Finally,
Appendix~\ref{sec:Vertical} provides further insight into vertical cuts at $ h = 4, 6$ and~$10$.

\section{Two limiting cases: ETH and Anderson}
\label{sec:warmup}
To set the stage for the discussion, we present the main features of the spin-spin correlations and their probability distributions, highlighting the differences between two well-controlled limits. We first consider the ``many-body'' Anderson-localized phase ($\Delta = 0$, Sec.~\ref{sec:AL}), before going deep into the ergodic phase ($\Delta = 1, h= 1$, Sec.~\ref{sec:ETH}), where the ETH holds. 
\subsection{The many-body Anderson insulator}
\label{sec:AL}

\subsubsection{Exponential decay of the correlation functions}

In the absence of interaction $\Delta=0$, the XXZ Hamitonian becomes equivalent (after Jordan-Wigner fermionization) to free fermions in a random potential (half-filling corresponding to $S^{z}_{\rm{tot}} = 0$). Using Wick's theorem, it is straightforward~\cite{lieb_two_1961,pfeuty_one-dimensional_1970} to express the various correlation functions in terms of the Anderson localized fermionic orbitals $\phi_m(i)$ (see Appendix~\ref{sec:fermions} for free-fermion calculations). For instance one easily sees that the longitudinal spin-spin correlations, equivalent to the density-density correlations for fermions, are always negative (for any $i\neq j$):
\bea
\label{eq:Czz1}
    C^{zz}_{ij}  &=& -\left|\langle c^\dagger_i c^{\vphantom{\dagger}}_{j} \rangle\right|^2
    =- \left| \sum_{m} \phi^*_{m}(i) \phi_{m}({j}) \right|^2,
\eea
where $m$ labels the occupied AL orbitals. In contrast, the transverse correlations can be positive or negative
\bea
    \label{eq:Cxx1}
   C^{xx}_{ij} &=& \frac{1}{2} \langle S_i^{+} S_{j}^{-} \rangle \\
   &=& \frac{1}{2} \langle c_i^{\dagger} (-1)^{\hat{\varphi}_{ij}} c_{j}^{\vphantom{\dagger}}\rangle\quad({\rm{with}}~ {\hat{\varphi}_{ij}}=\sum_{\ell = i}^{j-1}c^{\dagger}_\ell c^{\vphantom{\dagger}}_\ell).
    \label{eq:Cxx2}
    \eea

\begin{figure}[t!]
    \centering
    \includegraphics[width=\columnwidth]{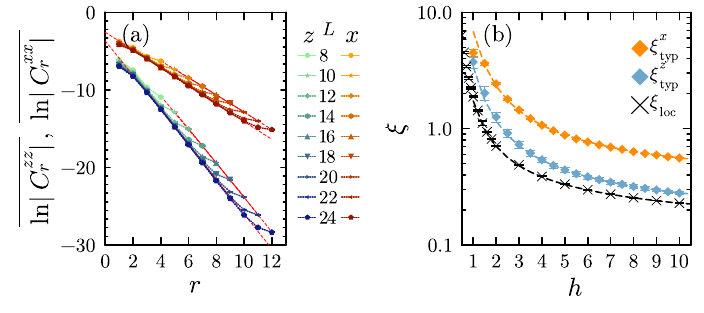}
    \caption{\label{fig:AL1} Typical correlations and associated lengths in the XX chain (Anderson localized insulator) at $\epsilon = 0.5$, averaging over $5 \cdot 10^4$ disorder realizations. Small sizes are taken in order to facilitate the comparison with the interacting model.
    (a) Exponential decay of the typical correlations at $h = 5$. Errors are smaller than the data points. The bulk correlation lengths $\xi^{z}_{{\rm bulk, typ}} = 0.448 \pm 0.005$ and $\xi^{x}_{{\rm bulk, typ}} = 0.868 \pm 0.006$  agree with the midchain correlation lengths $\xi^{z}_{\rm typ} = 0.449 \pm 0.003$ and $\xi^{x}_{\rm typ} = 0.894 \pm 0.009$.
    (b) Evolution of the midchain typical correlation length compared to the many-body localization length $\xi_{\rm loc}$ obtained from the Lyapunov exponent. 
    Dotted lines indicate the fits, Eqs.~\eqref{eq:xiloc},~\eqref{eq:xizxfit}. At weak disorder, finite-size effects limit the typical midchain correlation length.}
\end{figure}

In this non-interacting, Anderson-localized case, the correlations are expected to decay exponentially with the separation $r = |i-j|$, as shown by ED results in Fig.~\ref{fig:AL1} (a) for the typical correlations Eq.~\eqref{eq:typical}. We note that the rounding of $\overline{\ln|C^{\alpha\alpha}_{r}|}$ observed at large $r$ and fixed size $L$ is an effect of the periodic boundary conditions. Importantly, in this case the \emph{bulk} decay of the typical correlations for $1<r<L/2-2$:
\be
C^{\alpha\alpha}_{r, \mathrm{typ}} \propto \exp\left(-\frac{r}{\xi^\alpha_{\mathrm{bulk, typ}}}\right)
\ee
agrees with the decay of the \emph{midchain} typical correlations taken at $r=L/2$ as a function of system size $L$
\be
C^{\alpha\alpha}_{L/2, \mathrm{typ}} \propto \exp\left(-\frac{L}{2\xi^{\alpha}_{\mathrm{typ}}}\right).
\label{eq:CL2exp}
\ee

It is instructive to compare these correlation lengths with a many-body version of the Anderson localization length, which can be computed from the Lyapunov exponent associated with single-particle orbitals~\cite{muller_delande_notes,colbois_breaking_2023,crowley_avalanche_2020}. The single-particle localization length depends on both disorder strength and single-particle energy.  To obtain an overall localization length $\xi_{\mathrm{loc}}$ characteristic of \emph{many-body} eigenstates at a given many-body energy target (e.g. $\epsilon =0.5$~\footnote{Here $\epsilon=(E-E_{\rm min})/(E_{\rm max}-E_{\rm min})\in[0,\,1]$ defines the (many-body) energy density above the ground-state energy $E_{\rm min}$, with $E$ the total energy and $E_{\rm max}$ the maximal energy.}), we have to average over the density of states~\cite{colbois_breaking_2023,crowley_avalanche_2020}. Fig.~\ref{fig:AL1} (b) shows the disorder dependence  of $\xi^{\alpha}_{\rm typ}$ and $\xi_{\rm loc}$. Inspired by the fact that the latter is well described by the form~\cite{potter_universal_2015,colbois_breaking_2023} 
\be 
    \label{eq:xiloc}
    \xi_{\rm loc}(h) =  \frac{1}{\ln\left[1+(h/h_0)^2\right]}, \quad h_0 \approx 1.13
\ee    
we numerically fit the typical correlation lengths and obtain similar behaviors for $h \gtrsim 1.5$
\be
    \xi^{x}_{\rm{typ}}(h)\approx  2 \xi^{z}_{\rm{typ}}(h)\approx \frac{1}{\ln\left[1+(h/h'_0)^2\right]}, \quad h'_0 \approx 1.72.
   \label{eq:xizxfit}
\ee

The estimated midchain correlation lengths plotted in Fig.~\ref{fig:AL1} (b) deviate slightly from their expected behavior for $h \lesssim 1.5$. For $\xi^z_{\rm typ}$, this is a very simple finite-size effect due to the small sizes considered here, of the same order as those reachable in the interacting problem. We note that this occurs for $\xi^{z}_{\rm typ} \gtrsim 3$. For $\xi^{x}_{\rm typ}$, the finite-size effects are more subtle: the correlation length shows a maximum as a function of the disorder strength, drifting to zero disorder as the size increases. For stronger random fields, the fits are good and the ratio of the transverse to longitudinal correlation lengths is $\approx 2$.

There is a simple argument to explain the factor of 2: for strong enough disorder, the local densities are very close to 1 or 0, and therefore in most cases we expect the Jordan-Wigner phase factors ${\hat{\varphi}_{ij}}=\sum_{\ell = i}^{j-1}c^{\dagger}_\ell c^{\vphantom{\dagger}}_\ell$ to contribute almost exclusively to the overall sign in Eq.~\eqref{eq:Cxx2}, leading to 
\bea
\label{eq:Cxxappr1}
C^{xx}_{ij}&\approx& \pm \frac{1}{2}\langle c^\dagger_i c^{\vphantom{\dagger}}_{j} \rangle\\
&\approx& \pm \frac{1}{2} \sqrt{\left|C^{zz}_{ij}\right|}\label{eq:Cxxappr2},
\eea
and thus to the observed proportionality between the two components of the correlation lengths
\be
\xi^x_{\rm typ}\approx 2\xi^z_{\rm typ}.
\label{eq:xixz}
\ee

The \emph{average} midchain correlations also decay exponentially in the non-interacting case (not shown here, but see Sec.~\ref{sec:avg_typ}), yielding a midchain "average" correlation length: 
\begin{equation}
    \overline{|C^{\alpha\alpha}_{L/2}|} \propto \exp\left( -\frac{L}{2 \xi^{\alpha}_{\rm avg}}\right)
    \label{eq:Cavgmid}
\end{equation}
We note two important differences with the typical correlations stemming from the fact that the average correlations are dominated by the large values: the average is more noisy, and the longitudinal and transverse correlation lengths are not related by a simple constant ratio. Instead we observe that $\xi^{x}_{\rm avg} < 2 \xi^{z}_{\rm avg}$, as further discussed in Appendix~\ref{sec:fermions}.

\subsubsection{Histograms and correlation map between $C^{xx}_{L/2}$ and $C^{zz}_{L/2}$}

\begin{figure}[t!]
    \centering
    \includegraphics[width=\columnwidth]{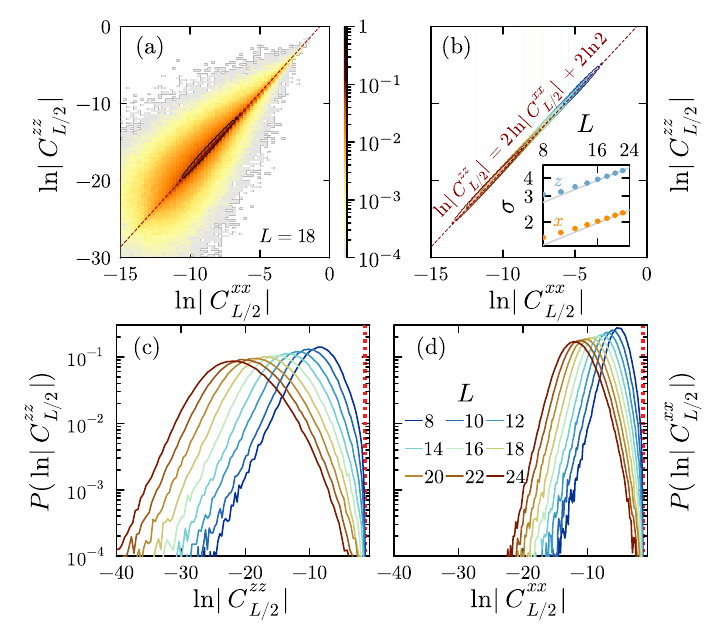}
    \caption{\label{fig:AL2} Distribution of the midchain correlations in the XX chain (Anderson localized) at $\epsilon = 0.5$, for $h = 3.5$, taken over $5 \cdot 10^4$ disorder realizations. The system size color code is the same in all panels.
    (a) Joint distribution for $L= 18$. The heatmap encodes $\mathcal{P}(\ln|C^{xx}|, \ln|C^{zz}|)$, in logarithmic scale.
    (b) Evolution with increasing system sizes of the joint distribution. For readability we only plot the contour at half the maximum probability (shown in black on panel (a))
    (c) and (d) : distribution of the natural logarithm of the longitudinal ($z$) and transverse ($x$) correlations. The bright red dotted lines indicate the maximal value of the correlations, $\ln|C^{\alpha\alpha}| \leq -\ln(4)$.
    The size dependence of the variance of the log distribution is shown in inset of panel (b), with the square root growth marked by a light gray line. }
\end{figure}

Keeping in mind our objective of exploring long-range correlations, we focus on the mid-chain correlation functions ($|i-j|=L/2$). In the panels (a) and (b) of Fig.~\ref{fig:AL2}, we consider the "correlations of the correlations" at disorder $h = 3.5$, exposing how transverse and longitudinal responses are correlated at the level of individual eigenstates. Most of the weight essentially accumulates along the red dashed line 
\be
    \ln|C^{zz}_{L/2}| = 2\ln|C^{xx}_{L/2}| +2\ln 2,
    \label{eq:XXZZAL}
\ee
in agreement with Eq.~\eqref{eq:Cxxappr2}. 
This behavior does not change with increasing system sizes, as shown in Fig.~\ref{fig:AL2}(b). Although Eq.~\eqref{eq:XXZZAL} is the typical behavior, which is not rigorously valid for each correlation at the individual eigenstate level, the results in Fig.~\ref{fig:AL2} (a-b) clearly reinforce the above arguments of the irrelevance of the Jordan-Wigner string, leading to Eqs.~(\ref{eq:Cxxappr1}) and (\ref{eq:Cxxappr2}), and thus the factor of two in the correlations lengths Eq.~\eqref{eq:xixz}. A simple physical interpretation is that the random magnetic field strongly pins and constrains the spins along the field direction (whether aligned with or against the field), and the dominant quantum fluctuations occur in the transverse channel, perpendicular to the random field direction.

When studied independently, the individual distributions of the $xx$ and $zz$ midchain correlations, plotted in Fig.~\ref{fig:AL2}(c,d) for $h=3.5,\, \Delta = 0$, show very interesting features as system size is increased. Here we focus on small lengths ($L=8,\,\ldots,\,24$) to favor the comparison with the interacting problem later on. 
Except for the already discussed factor of two, both families of distributions behave similarly: 

{\it (i)} They are upper bounded by $\ln|C^{\alpha\alpha}_{L/2}|\leq - \ln(4)$.

{\it (ii)} They drift toward negative values linearly with $L$ as ${\overline{\ln |C^{\alpha\alpha}_{L/2}|}}\sim -L/2\xi^{\alpha}$.

{\it (iii)} Their standard deviation increases, with a slower power-law compatible with 
\be
\sigma_{\ln |C^{\alpha\alpha}_{L/2}|}\sim \sqrt{L},
\label{eq:sigmalogCAL}
\ee
as shown in the inset of Fig.~\ref{fig:AL2}(b) (see also Appendix~\ref{sec:fermions}). We note that the behavior Eq.~\eqref{eq:sigmalogCAL} and the presence of logarithmic corrections has been conjectured by Fisher and Young~\cite{fisher_distributions_1998} for the related many-body AL regime of the random transverse-field Ising chain. These distributions are not perfectly well captured by (even truncated) simple log-normal distributions~\footnote{The large correlations are captured correctly but not the weak-correlations tail.}.

\subsection{Deep in the ergodic phase}
\label{sec:ETH}

\begin{figure*}[t!]
    \includegraphics[width=\textwidth]{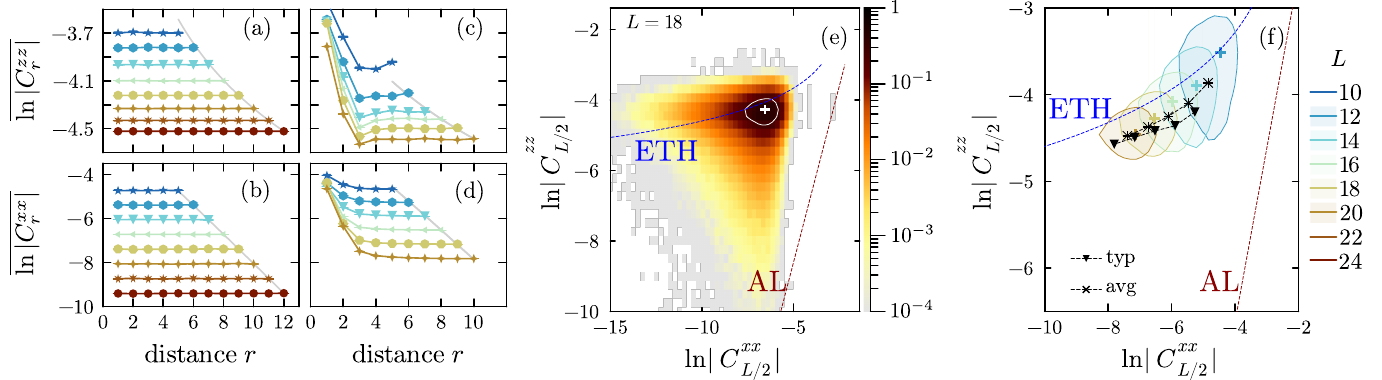}
    \caption{\label{fig:Ergodic1}Typical decay of the spin-spin correlations of a random vector in the $S^{z}_{\rm{tot}} = 0$ sector compared to results deep in the ergodic phase at $\Delta = 1, h = 1$. 
    (a) Longitudinal corrrelations for the random vector. The gray line corresponds exactly to Eq.~\eqref{eq:Czzdecay}.
    (b) Transverse correlations in the random vector. The gray line is given by Eq.~\eqref{eq:Cxxrandomdecay}
    (c) Longitudinal correlations in the ergodic phase. The gray line is given by Eq.~\eqref{eq:Czzactualdecay}.
    (d) Transverse correlations in the ergodic phase. The gray line is a fit to a square-root decay with the $S^{z}_{\rm tot} = 0$ sector's size.
    (e) Joint distribution of the midchain correlations for $L = 18$, with the white contour indicating the height at half-maximum and the plus sign indicating the locations of the maxima.
    (f) Evolution of the contour at half-maximum of the joint midchain correlations distribution with increasing system sizes. The colored plus signs indicate the locations of the maxima of the distribution. Typical and average values of $|C^{\alpha \alpha}_{L/2}|$ are shown by triangles and crosses respectively. In panels (e) and (f), the AL line is given by Eq.~\eqref{eq:XXZZAL} while the ETH line is given by Eqs.~\eqref{eq:Cxxrandomdecay} and~\eqref{eq:Czzdecay}.}
\end{figure*}

\subsubsection{ETH and random states}

In the ergodic phase, a typical ``infinite temperature'' eigenstate (in the middle of the many-body spectrum, $\epsilon=0.5$) should be well described by a featureless random state, with no spatial dependence of two-point correlations, namely $C_{ij}^{\alpha\alpha}$ is independent of $|i-j|$. The finite-size effects in expectation values should be entirely ascribed to the finite-dimension ${\cal N}_{\cal H}$ of the available Hilbert space, and for some quantities the existence of an underlying sum rule due to a conservation law. In the remainder of this paper, the size of the available Hilbert space ${\cal N}_{\cal H}$ will be the sizes of the $S^{z}_{\rm tot} = 0$ magnetization sector. This leads to markedly different behavior as a function of system size $L$ for longitudinal and transverse correlations, clearly visible in the two leftmost panels (a,\,b) of Fig.~\ref{fig:Ergodic1}, which we now discuss.

\paragraph{Longitudinal correlations---} The conservation of the total magnetization $\sum_{i=1}^{L}S_i^{z}=S^z_{\rm{tot}}$ can be rewritten as follows
\be
    \sum_{i,j} \langle S_i^{z} S_j^{z} \rangle = \frac{L}{4} + \sum_{i \neq j} \langle S_i^{z} S_j^{z} \rangle=(S^z_{\rm{tot}})^2.
    \ee
For the $S^z_{\rm{tot}}=0$ sector (for even $L$), when associated to the featureless property of a random vector, one expects on average
    \be
    \overline{\langle S_i^z S_j^z \rangle} \approx  -\frac{1}{4(L-1)} \quad \forall i\neq j.
    \label{eq:Czzdecay}
\ee
While Eq.~\eqref{eq:Czzdecay} applies to the average disconnected correlations, it is also valid for the typical, connected correlations, up to corrections decaying exponentially with system size. The absence of dependence on $r$ and the decay of the typical midchain correlations with $L$ is readily checked in Fig.~\ref{fig:Ergodic1}(a) against exact numerics for fully random states in the $S^{z}_{\rm tot} = 0$ sector, averaged over hundreds of samples~\footnote{More than 1500 samples for $L = 18$ and below, and more than 600 for $20 \leq L \leq 24$}.

\paragraph{Transverse correlations---} 
In contrast to the above slow algebraic decay, the transverse component is suppressed much faster with $L$. Since  for a random state the correlations do not depend on the distance, $\langle S_i^+ S_j^-\rangle$ is given by a sum of ${\mathcal{N}}_{\cal H}/4$ variables of order $1/{{\mathcal{N}}_{\cal H}}$ with random signs, where ${\mathcal{N}}_{\cal H}$ is the size of the Hilbert space. We therefore expect
\begin{equation}
    \overline{\langle S_i^x S_j^x\rangle}  \propto \frac{1}{\sqrt{{\mathcal{N}}_{\cal H}}} \quad \forall i\neq j,
    \label{eq:Cxxdecay}
\end{equation}
a behavior that is checked for $S^{z}_{\rm tot} = 0$ random vectors in Fig.~\ref{fig:Ergodic1}(b), where we observe
\be
    C^{xx}_{L/2, {\rm typ}} \approx \frac{0.137}{\sqrt{\mathcal{N}_{\cal H}}},
    \label{eq:Cxxrandomdecay}
\ee
with ${\mathcal{N}_{\cal H}}= \frac{{L!}}{{(L/2)!^2}}\sim 2^L/\sqrt{L}$ for the largest ($S^{z}_{\rm{tot}} = 0$) sector.

\subsubsection{ED results deep in the ergodic regime}

\begin{figure}[b!]
	\centering
	\includegraphics[width=\columnwidth]{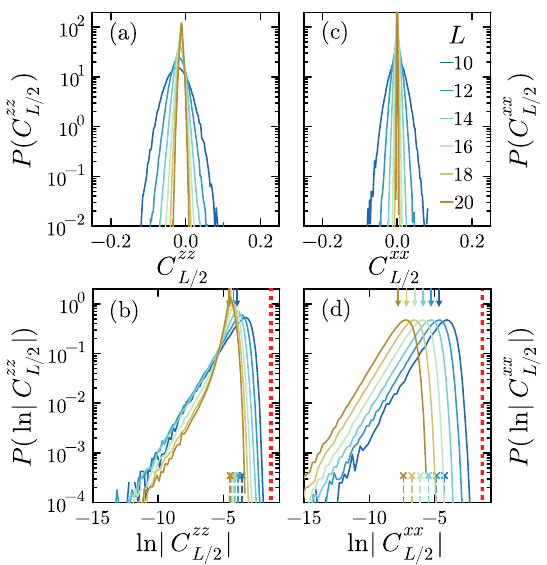}
	\caption{\label{fig:Ergodic2} Distributions of the midchain spin-spin correlations deep in the ergodic phase, at $\Delta = 1, h=1$. (a)-(b) Longitudinal correlations. 
    The majority of the correlations are negative. 
    The variance of the distribution of the logarithm clearly decreases with increasing system sizes.
    (c)-(d) Transverse correlations. The distributions of $|C^{xx}_{L/2}|$ are gaussian and result in a traveling-wave behavior for the distribution of $\ln|C^{xx}_{L/2}|$. 
    In the bottom panels, the bright red dotted lines indicate the maximal value of the correlations, $\ln|C^{\alpha\alpha}| \leq -\ln(4)$. The arrows indicate the typical value of $|C^{\alpha \alpha}_{L/2}|$, while the crosses indicate the average. 
    }
\end{figure}

We now discuss the more realistic situation of the XXZ chain with parameters $h=1, \Delta=1$, i.e. deep in the ergodic regime, where we expect the above ETH behavior to apply. Figs.~\ref{fig:Ergodic1} (c,\, d) show the relatively good agreement with the analytical forms Eq.~\eqref{eq:Czzdecay} and Eq.~\eqref{eq:Cxxdecay}, despite some small discrepancies at short distance. 

In order to minimize this non-universal residual effect, we focus on midchain correlations, with a leading term that indeed scales as a power-law for $zz$, shown by a gray line in Fig.~\ref{fig:Ergodic1}(c):
\be
    C^{zz}_{L/2,\typ} \approx \frac{0.27}{L + \ell_0},
    \label{eq:Czzactualdecay}
\ee
where the shift $\ell_0\approx 6.2$ is due to the small size dependence at short distance.
The transverse ($xx$) typical correlator is better described by the random vector results, the gray line corresponding to $C^{xx}_{L/2, \typ}  =  0.17 /\sqrt{{\mathcal{N}}_{\cal H}}$ in Fig.~\ref{fig:Ergodic1}(d)

In order to emphasize the contrasting behavior of the correlators in the ergodic and Anderson-localized phase, we show in a heatmap the joint probability distribution $\mathcal{P}(\ln|C^{xx}|, \ln|C^{zz}|)$ for $L=18$ in Fig.~\ref{fig:Ergodic1} (e), and the size dependence of its half-maximum in Fig.~\ref{fig:Ergodic1} (f), which can be compared with Fig.~\ref{fig:AL2} (a,\,b) for the AL case. Clearly, the distribution of correlations between the two components $C_{L/2}^{xx}$ and $C_{L/2}^{zz}$ take a different (triangular-like) shape, with a much smaller variance rapidly decaying with system size. The typical, average and maximal value rapidly converge towards each other, and follow roughly the ETH behavior (from random states, displayed in  blue, long-dashed lines in the Figure), up to the correction due to $\ell_0$.
This behavior contrasts with the convergence towards the AL line ($\ln|C^{zz}_{L/2}| = 2\ln|C^{xx}_{L/2}| +2\ln 2$), see Fig.~\ref{fig:AL2} (a,b), and the red dashed line in Fig.~\ref{fig:Ergodic1} (f).

In addition, we also show in Fig.~\ref{fig:Ergodic2} the individual distributions $P(C_{L/2}^{\alpha\alpha})$ (panels a,c) and $P(\ln|C_{L/2}^{\alpha\alpha}|)$ (panels b, d). They can be compared with the AL case, displayed in Fig.~\ref{fig:AL2} (c-d), and with random states (see Fig.~\ref{fig:RandomVector} in Appendix~\ref{sec:RS}).
This set of plots showcases not only the strong differences between AL and ergodic behaviors, but also between the two different components (transverse and longitudinal) inside the ergodic phase. In the top panels of Fig.~\ref{fig:Ergodic2}, one immediately remarks the strong sharpening with system size of the peaks of $P(C^{\alpha \alpha}_{L/2})$ around their mean values. This is typical of the self-averaging nature of the ergodic phase. More precisely we find (see Appendix~\ref{sec:variance}) that the standard deviation of $C^{\alpha \alpha}_{L/2}$ decreases rapidly with the Hilbert space size as 
\be
\sigma_{C_{L/2}^{\alpha\alpha}}\sim \frac{1}{\sqrt{{\cal N}_{\cal H}}}.
\label{eq:sigmaETH}
\ee
Furthermore, in agreement with the random state predictions of Eq.~\eqref{eq:Czzdecay}, we find that the average longitudinal correlation $\overline{C^{zz}_{L/2}}$ takes mostly negative values, in contrast with the transverse correlation $C^{xx}_{L/2}$ which has a symmetric distribution around zero.  

The lower panels of Fig.~\ref{fig:Ergodic2} show the probability distributions of the logarithm of the correlators $P(\ln|C^{\alpha\alpha}_{L/2}|)$ where the difference between the longitudinal and transverse components is much more visible. In particular, the standard deviation for the $zz$ component decreases very fast 
\be
\sigma_{\ln |C_{L/2}^{zz}|}\sim \frac{L}{\sqrt{{\cal N}_{\cal H}}},
\ee
as a simple consequence of ETH Eq.~\eqref{eq:sigmaETH} and the conservation of the total magnetization yielding Eq.~\eqref{eq:Czzdecay}. In contrast, $P(\ln|C^{xx}_{L/2}|)$ shows a clear traveling wave behavior with a constant standard deviation
\be
\sigma_{\ln |C_{L/2}^{xx}|}\sim {\cal{O}}(1),
\ee
which is also a consequence of ETH, since both $\sigma_{C_{L/2}^{xx}}$ and $C_{L/2}^{xx}$ decay $\sim 1/\sqrt{\cal N_H}$. 

\subsection{Summary of the two limits}
To summarize the main results in the two limiting cases (many-body AL and deep ergodic case, as captured by random states), we report in Table ~\ref{tab:1} the different scaling behaviors as a function of chain length $L$ or Hilbert space size ${\cal N}_{\cal H}$ for midchain correlators, related correlation lengths, and variance of probability distribution. We find that the midchain correlators (at $r=L/2$) are particularly useful for several reasons: (i) in the ETH phase, they naturally capture the strongly different decays (exponential versus algebraic) of the $C^{xx}$ and $C^{zz}$ correlators; (ii) in this same phase, they are less affected by finite-size effects that influence short-distance correlators (at small $r$); (iii) in the AL phase they perfectly capture the correlation lengths $\xi^x$ and $\xi^z$, equally well if not better than bulk correlators. 

\begin{table}[t!]
\begin{center}
\begin{tabular}{c||c|c}
 & ETH & Many-body AL\\
\hline
&&\\
Mid-Chain& ${\overline{\ln |C_{L/2}^{xx}|}}\sim -\ln{\sqrt{{\cal N}_{\cal H}}}$ & ${\overline{\ln |C_{L/2}^{xx}|}}\sim -\frac{L}{2\xi^x}$ \\
Correlations& &\\
Functions&${\overline{C_{L/2}^{zz}}}\approx -\frac{1}{4(L-1)}$ & ${\overline{\ln |C_{L/2}^{zz}|}}\sim -\frac{L}{2\xi^z}$\\
&&\\
\hline
&&\\
Correlation &&\\
lengths & $~\xi^x_{\rm typ}\sim {\cal{O}}(1),\, \xi^{z,\rm eff}_{\rm typ}\to\infty~$ & $\xi^x_{\rm typ}=2\xi^z_{\rm typ}\sim {\cal{O}}(1)$\\
&&\\
\hline
&&\\
Standard&$\sigma_{\ln |C_{L/2}^{xx}|}\sim {\cal{O}}(1)$& $\sigma_{\ln |C_{L/2}^{xx}|}\sim \sqrt{L}$\\
Deviations& &\\
& $\sigma_{\ln |C_{L/2}^{zz}|}\sim \frac{L}{\sqrt{{\cal N}_{\cal H}}}$ & $\sigma_{\ln |C_{L/2}^{zz}|}\sim \sqrt{L}$\\
      \bottomrule
\end{tabular}
\caption{Summary of the different behaviors for the transverse ($xx$) and longitudinal ($zz$) components of the midchain correlation functions $C_{L/2}^{\alpha\alpha}$ for the ergodic (ETH) and Anderson localized (AL) limits. ${\cal N}_{\cal H}$ is the dimension of the $S^{z}_{\rm tot} = 0 $ magnetization sector.
\label{tab:1}}
\end{center}
\end{table}

\begin{figure*}
	\centering
\includegraphics[width=\textwidth]{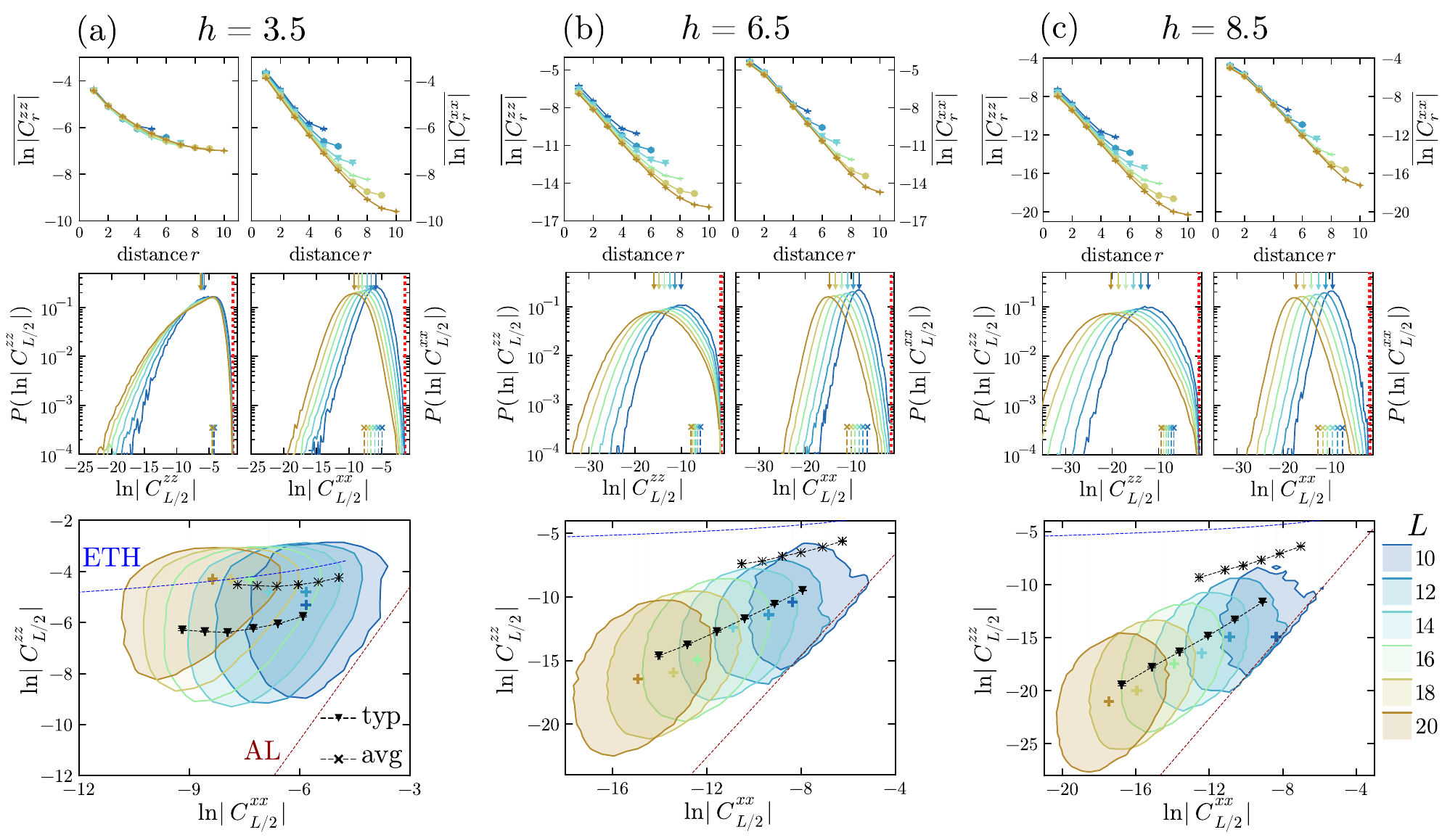}
	\caption{\label{fig:overview} Overview of the spin-spin correlations along the RFHM line ($\Delta=1$) for three representative disorder strengths (a)~$h=3.5$, (b)~$h = 6.5$ and (c)~$h = 8.5$, for even system sizes from $L = 10$ to $L = 20$.  
    {First row:} Decay of the typical longitudinal (left) and transverse (right) correlations with the distance $r = |i-j|$.
    {Second row:} Probability distribution $P(\ln |C^{\alpha \alpha}_{L/2}|)$. The short vertical lines denote horizontally the typical (triangles) and average (crosses) values of $|C^{\alpha \alpha}_{L/2}|$. The bright red dotted lines indicate the maximal value of the correlations, $\ln|C^{\alpha\alpha}_{L/2}| \leq -\ln(4)$. {Third row:} Evolution of the contour at half-maximum of the joint distribution ${P}(\ln |C^{xx}_{L/2}|,\ln |C^{xx}_{L/2}|)$ with increasing system sizes. The colored plus signs indicate the locations of the maxima of the distribution. Typical and average values of $|C^{\alpha \alpha}_{L/2}|$ are shown by triangles and crosses respectively.  For all disorder strengths, the typical and average values are bounded by the two asymptotic AL and ETH behaviors. The former is represented by a dashed red line corresponding to Eq.~\eqref{eq:XXZZAL}, and the latter by a dashed blue line given by Eqs.~\eqref{eq:Cxxrandomdecay} and~\eqref{eq:Czzdecay}.}
\end{figure*}

In the next sections, we explore in detail how the correlations and associated correlation lengths evolve as a function of interaction and disorder strengths. In Ref.~\cite{colbois_interaction_2024}, we described as a smoking gun the fact that the AL hierarchy $\xi^x_{\rm typ} \approx 2 \xi^z_{\rm typ}$ evolves to eventually switch, such that $\xi^x_{\rm typ} \approx   1/\ln 2  \ll \xi^{z,\mathrm{eff}}_{\rm typ} \to \infty$ in the ergodic phase. Here, our discussion goes much deeper into the distributions, the statistics of rare events,  and provides further insight by exploring the subtle differences between typical and average correlations.

\section{Correlations across the phase diagram}
\label{sec:Overview}
\subsection{Overview}
We now consider the fate of both transverse and longitudinal correlation functions over the entire XXZ phase diagram shown in Fig.~\ref{fig:phasediag}. This is first illustrated  in Fig.~\ref{fig:overview} with a horizontal cut performed at $\Delta=1$ (the RFHM) for three representative disorder strengths: one point in the ergodic regime ($h=3.5$) and two points of stronger disorder,  presumably belonging to the MBL phase ($h=6.5$ and $h=8.5$), at least according to the extrapolation of standard estimates~\cite{sierant_polynomially_2020,colbois_interaction_2024}. 
\subsubsection{Ergodic regime ($h=3.5$)}
We begin with a discussion of Fig.~\ref{fig:overview} (a), which shows spin-spin correlation data for $h=3.5$. 
Although less deep in the ergodic regime than the previous case ($h=1$ in Fig.~\ref{fig:Ergodic1} and Fig.~\ref{fig:Ergodic2}), these data are consistent with  ergodic behavior, and nicely show a crossover to the ETH limit. This is best seen in the lower panel, where the contour at the half-maximum of the joint distribution of the $xx$ and $zz$ midchain correlators ${P}(\ln|C^{xx}_{L/2}|,\ln|C^{zz}_{L/2}|)$ clearly evolves toward ETH behavior with increasing system size. 
This trend can also be observed in the typical correlation functions (top panels), where the clear exponential decay of the transverse component $C^{xx}(r)$ differs from the transient behavior of $C^{zz}(r)$, which signals a slow change towards ETH.

The histograms of the midchain correlations $P(\ln|C^{\alpha\alpha}_{L/2}|)$, shown in the middle panels, reveal additional features compared to $h=1$ in Fig.~\ref{fig:Ergodic2} (c-d). Indeed, the contrasted behaviors of the two components are different. Both distributions broaden with increasing system sizes, while at the same time an increasingly pronounced peak slowly develops for the $z$ component near the average value, signaling the onset of the ETH behavior.

Furthermore, typical ${\overline{\ln|C^{\alpha\alpha}_{L/2}|}}$ and average $\ln{\overline{|C^{\alpha\alpha}_{L/2}|}}$ correlations are indicated by down arrows and crosses on the histogram (middle panels), as well as in the $x-z$ correlation map (lower panel). For $h=3.5$, they show similar trends towards ETH as $L$ increases. The distinction between typical and average behaviors will be discussed below  in Sec.~\ref{sec:avg_typ}.

\subsubsection{Strong disorder ($h=6.5,\, 8.5$)}
For stronger disorder, the situation differs both qualitatively and quantitatively from the ergodic regime, as shown in Fig.~\ref{fig:overview} (b,\,c). In the upper panels, typical midchain correlations decay exponentially for both components (discussed further in Sec.~\ref{sec:avg_typ}), although $C^{zz}_r$ shows more rounding in the bulk.
Furthermore, the histograms of midchain correlations show a very different behavior compared to both ergodic and AL regimes. While the transverse correlations change only slightly with $h$, the longitudinal component $C_{L/2}^{zz}$ exhibits several remarkable properties.

{\it (i)} Together with a shift towards large negative values, a clear broadening is observed with increasing $L$ (middle panels)~\footnote{One can notice in Fig.~\ref{fig:overview} for $h=8.5$ and $L=20$ the weakest correlations start to reach the numerical precision when $\ln |C| < -36$)}. However, unlike the AL results in Fig.~\ref{fig:AL2} (c), a strong asymmetry develops with an accumulation of the weight in the right tail (large correlations). This effect is most pronounced at $h=6.5$.

{\it (ii)} The joint correlations distributions (bottom panels) are very different from the AL case (Fig.~\ref{fig:AL2} (b)). The contours at half maximum encompass a much broader area instead of accumulating on the line given by the typical decay. 

{\it (iii)} The bottom panels show that average and typical behaviors are different at both quantitative and qualitative levels. This is particularly noticeable at $h=6.5$, where there is a curvature for the average correlations, but not for the typical.

We explore this last and very significant effect in more detail in the next section, before further discussing important features of the distributions in Sec.~\ref{sec:distributions}.  

\subsection{Average {\it{vs.}} typical correlations}
\label{sec:avg_typ}
We now address the distinction between average and typical correlations along two representative scans across the phase diagram, namely varying disorder $h$ with fixed interaction $\Delta=1$ (RFHM), and then varying interaction $\Delta$ for a fixed disorder strength $h=6$. Other cuts are shown in Appendix~\ref{sec:Horizonal} ($\Delta = 0.25, 0.5, 0.75$) and~\ref{sec:Vertical} ($h=4$, and more details on $h=6$).

\subsubsection{Midchain correlation functions}
\begin{figure}[t!]
\centering
\includegraphics[width=\columnwidth]{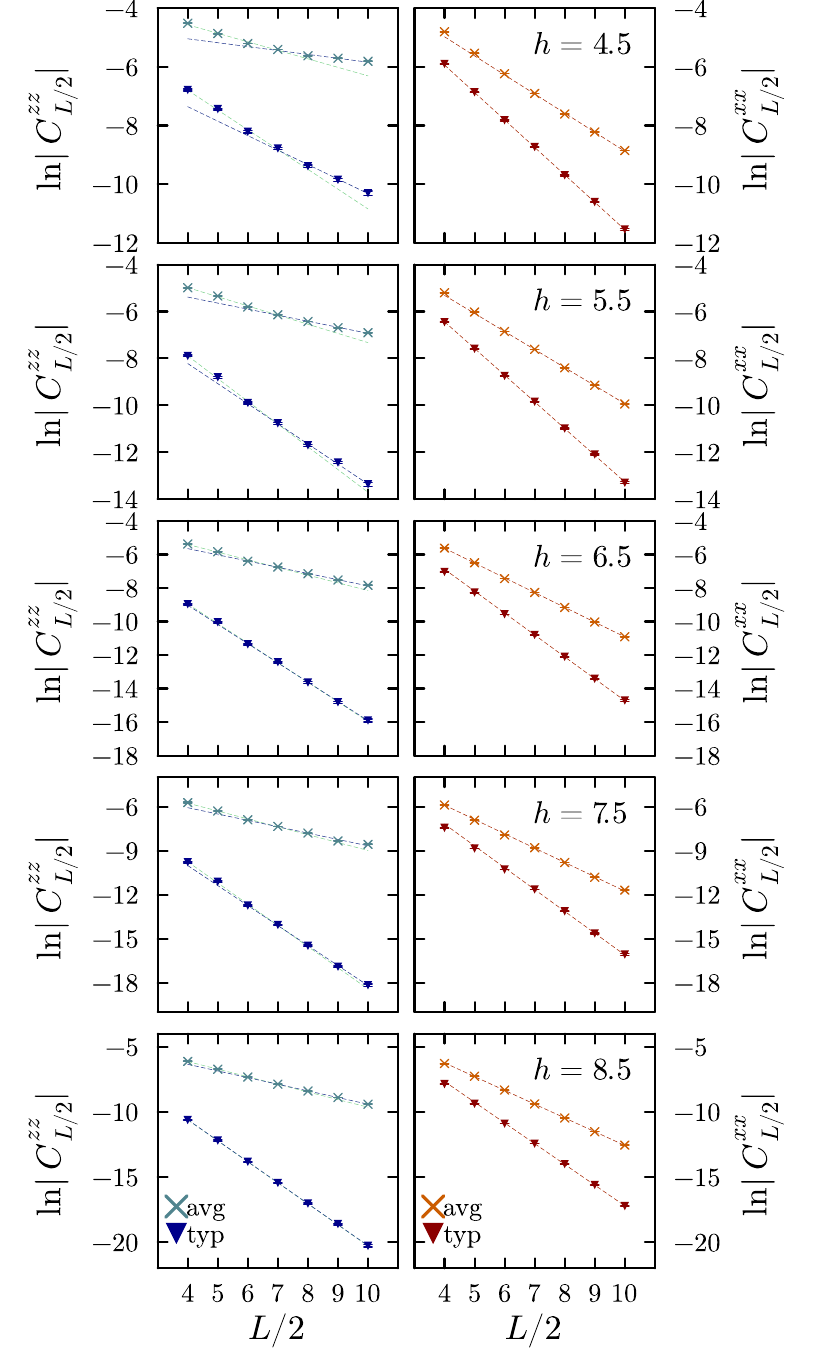}
\caption{\label{fig:typavgD1}
Evolution of longitudinal ($zz$, left) and transverse ($xx$, right) midchain correlations. The typical $\overline{\ln |C^{\alpha \alpha}_{L/2}|}$ (triangles) and average $\ln \overline{|C^{\alpha \alpha}_{L/2}|}$ (crosses) are plotted against $L$ for $\Delta=1$ and various disorder strengths (shown on plots). A linear scale is used to highlight possible exponential decays. Dashed lines show exponential fits. For $zz$ (left), two fits are shown: light green for the four smallest sizes, and dark blue for the four largest, qualitatively revealing fitting instabilities. For $xx$ (right), we only show one fit.}
\end{figure}

\begin{figure}[t!]
\centering
\includegraphics[width=\columnwidth]{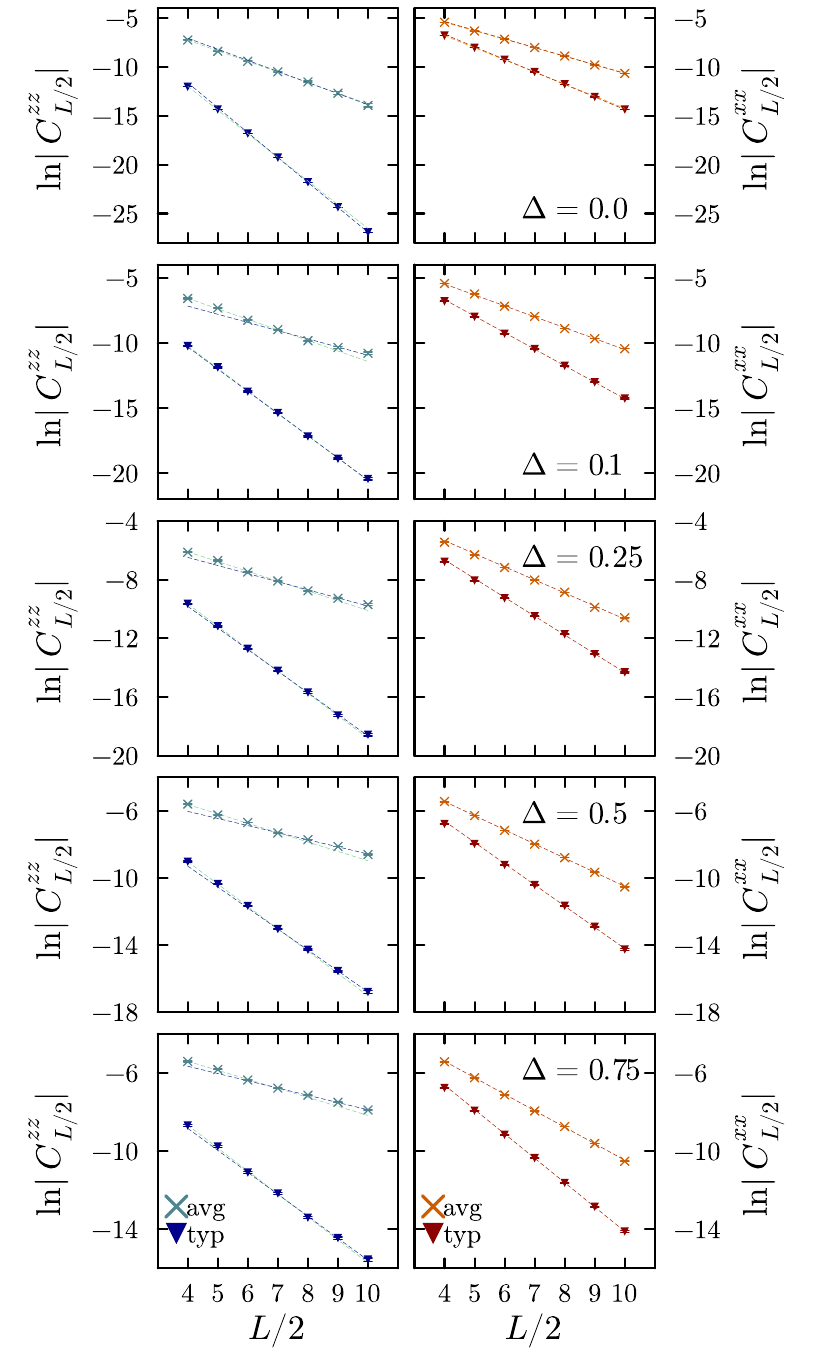}
\caption{\label{fig:typavgh6} Same as for Fig.~\ref{fig:typavgD1}, but for a fixed value of disorder $h=6$ and various values of interaction (from top to bottom). The first panel corresponds to the non-interacting many-body AL insulator.}
\end{figure}

\paragraph{$\Delta=1$: horizontal scan---}Fig.~\ref{fig:typavgD1} displays along the RFHM line the evolution with increasing disorder strength ($h=4.5,\,\ldots,\,8.5$) of the average $\ln \overline{|C^{\alpha \alpha}_{L/2}|}$ and the typical $ \overline {\ln {|C^{\alpha \alpha}_{L/2}|}}$ midchain correlations for both longitudinal ($\alpha = z$) and transverse ($\alpha = x$) components. As already noticed, the transverse correlations remain very well described by a purely exponential decay,  with the additional observation that average correlations decay with a larger correlation length: $C_{L/2}^{xx}$ is not sensitive to the transition, and behaves essentially the same in all regimes. 

In sharp contrast, the longitudinal correlations $C_{L/2}^{zz}$ show very different properties as $h$ varies. 
Starting from strong  disorder, we first find that average and  typical correlations appear to decrease exponentially over the whole range of $L$ for $h \gtrsim 8$. For $8 \gtrsim h \gtrsim 6$, some deviations start to become very visible for the average which no longer follows a pure exponential decay over the full range. Indeed, exponential fits (dashed lines) do not display the same slope depending on the fitting range in the left panels of Fig.~\ref{fig:typavgD1}. 
The discrepancy becomes increasingly important as $h$ decreases, while at the same time the typical correlations instead seem to remain mostly better described by an exponential decay, at least for $h\ge 6$, where the various fits agree reasonably well. 
A more systematic analysis in terms of the effective longitudinal correlation length (dependent on the fitting window) is given below in Sec.~\ref{sec:corrl}.

\paragraph{$h=6$: vertical scan---} The results along a vertical cut also display interesting features, as shown in Fig.~\ref{fig:typavgh6} for increasing interaction ($\Delta=0,\,\ldots,\, 0.75$) at $h=6$. As already observed in Ref.~\cite{colbois_interaction_2024}, for such a disorder the transverse correlations are practically insensitive to the interaction strength, while the longitudinal component is remarkably enhanced by any $\Delta\neq 0$: this is easily seen by comparing the left and right panels of Fig.~\ref{fig:typavgh6}.  Moreover, the distinction between average and typical values is particularly relevant for the $zz$ component. Indeed, as can be seen in the left panels, there is a striking contrast between $\Delta=0$ (top panel) and an interaction as weak as $\Delta=0.1$, where the average does not decay as a pure exponential over the whole range, while the typical value does. If we increase $\Delta$ further, the situation remains qualitatively the same, as we now analyze quantitatively by focusing on the correlation lengths.

\subsubsection{Correlation lengths}
\label{sec:corrl}

We first observe that for midchain transverse correlation functions $C_{L/2}^{xx}$,  the associated transverse correlation length $\xi^x$ is practically featureless, as it remains ${\cal{O}}(1)$ over the entire parameter space. The lengths $\xi^{x}_{\rm avg/typ}$ are very stable with system size, and barely depend on the averaging process.
This reinforces the result that the midchain transverse correlations  always decay exponentially with $L$ over the entire phase diagram. However, it is important to realize that the origin of these exponential decays is very different in the ergodic regime, where it comes from the simple ETH argument, see Eq.~\eqref{eq:Cxxdecay}, while at strong disorder it is a signature of localization.

\begin{figure}
\centering
\includegraphics[width=\columnwidth]{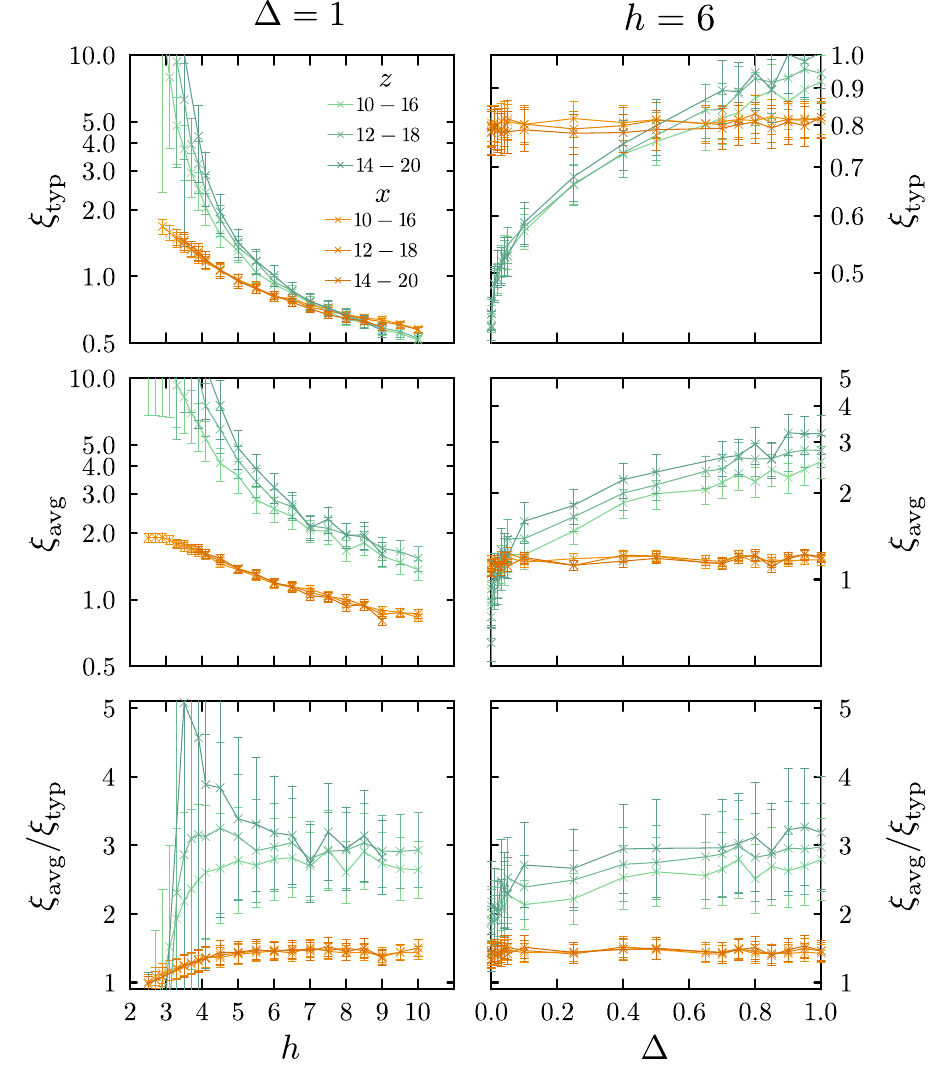}
\caption{\label{fig:xis} Correlation lengths associated to the typical (top) and average (middle) decays, and their ratio (bottom) in the longitudinal ($zz$, green) and transverse ($xx$, orange) directions. Left: $h$ dependence  at fixed interaction $\Delta =1$. Right: $\Delta$ dependence at $h=6$. Labels denote the various system sizes $L$ used for exponential fits to extract correlation lengths. The top panels were already presented in Ref.~\cite{colbois_interaction_2024}.}
\end{figure}

\begin{figure*}
    \centering
    \includegraphics[width=1.7\columnwidth]{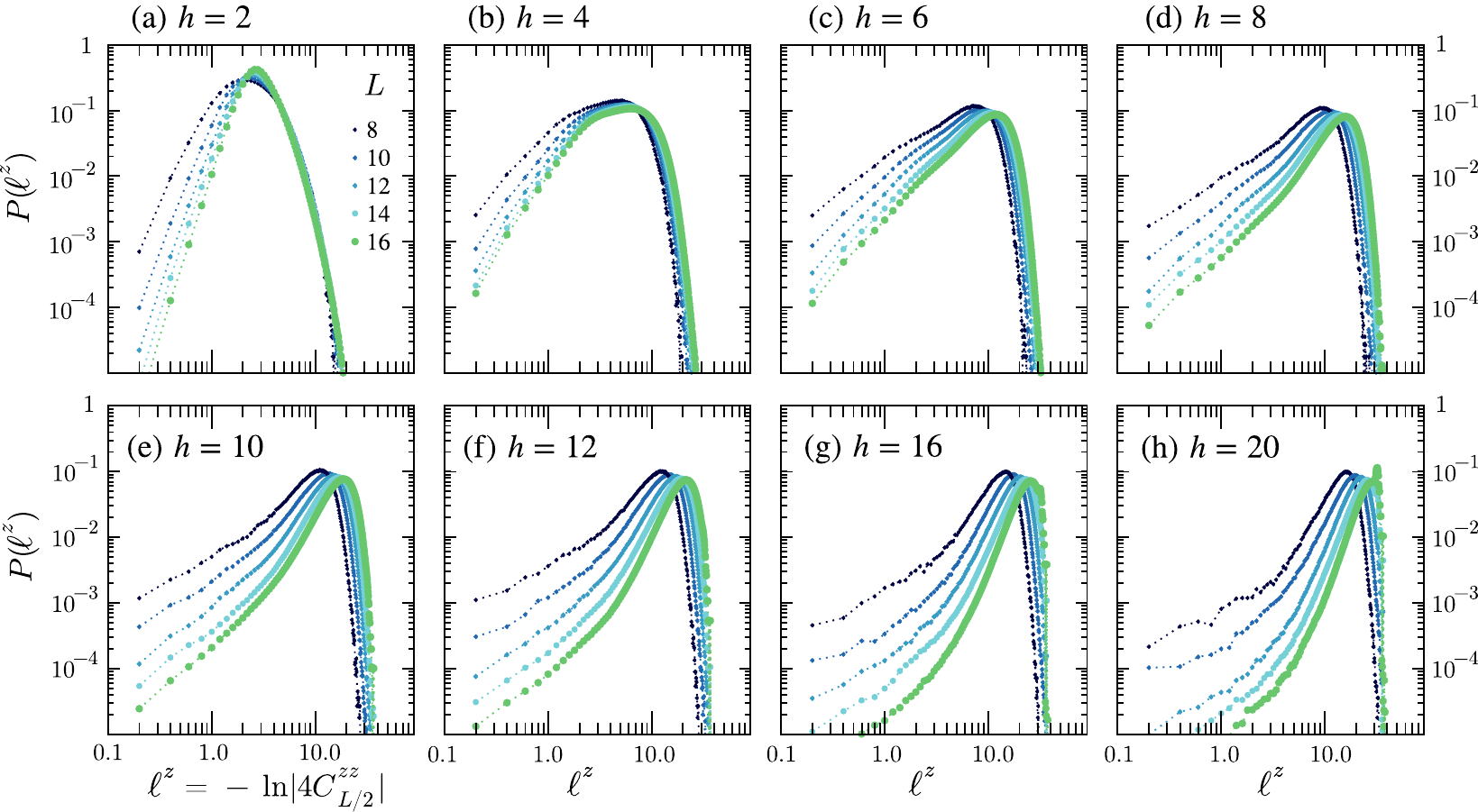}
    \caption{\label{fig:distrib_overview} Evolution of the distributions  $P(\ell^z)$ along the RFHM line ($\Delta=1$) for a wide range of disorder strengths $h\in [2,\,20]$. 
    The histograms, collected across {\emph{all}} eigenstates (from full diagonalization) and several thousands of disordered samples, are shown for various chain lengths $L=8,\,10,\,12,\,14,\,16$. As the disorder strength increases, two clear tendencies are observed: there is a first broadening, together with the development of a left fat tail visible for $h$ approximately in the range $[ 4,10]$, followed by a gradual vanishing of the  fat tail, which then becomes more and more Gaussian-like  as $L$ increases. For the largest values of $h$, a peak can be seen at the right-hand limit of the support ($\ell^z\approx 35$), corresponding to vanishingly small correlations that have reached machine precision.} 
\end{figure*}

The longitudinal length turns out to be much more sensitive to the parameters $\Delta$ and $h$, the system size $L$, and the averaging procedure. This can be clearly observed for example at $\Delta=1$ (Fig.~\ref{fig:xis} left), where we see that as $h$ decreases, $\xi^{z}_{\rm typ}$ starts to increase with $L$ when $h\le 5-6$, while $\xi^{z}_{\rm avg}$ also shows instability with increasing $L$, but already at $h\le 7-8$. A similar effect occurs at $h=6$ (Fig.~\ref{fig:xis} right) as a function of the interaction $\Delta$, where one sees that the enhancement with $L$ is much more pronounced for the average length than for the typical. 
The two lower panels of Fig.~\ref{fig:xis} provide another clear view of the strong difference between the average and typical correlation lengths, by showing how their ratio evolves. Despite the rather large error bars (due to the fact that we are taking a ratio of two estimates already averaged over the disorder), we can easily see that the ratio increases with $L$ below $h\approx 7$ for $\Delta=1$ and above $\Delta\approx 0.1$ for $h=6$, which signals a growing discrepancy between average and typical behavior. 

All these observations reinforce the idea that there might be a regime where the decrease of the average correlations is not well captured by a stable exponential decay. This calls for a closer inspection of the distributions of the longitudinal correlators $C_{L/2}^{zz}$, focusing in particular on certain rare events with large values.

\section{Distributions of correlations, extreme statistics, rare events}
 \label{sec:LargeCorrelations} 
\subsection{Focus on large correlations: general overview}
\label{sec:distributions}

To emphasize the large values of correlators, we consider for both components ($\alpha=x,\,z$) the strictly positive quantities
\be
\ell^{\alpha}=-\ln\left|4C_{L/2}^{\alpha\alpha}\right|,
\ee
and examine their distributions. Compared to the histograms $P(\ln|C_{L/2}^{\alpha\alpha}|)$ plotted in Fig.~\ref{fig:overview}, there is a clear advantage in showing $P(\ell^\alpha)$ instead as we directly focus on the \emph{large} correlation events $|C_{L/2}^{\alpha\alpha}|\approx 1/4$ corresponding to the ${\ell}^{\alpha}\to 0$ regime.

{\it Longitudinal correlators --- } Fig.~\ref{fig:distrib_overview} gives a broad overview of the evolution of $P(\ell^z)$ for different sizes $L=8 \cdots 16$ along the RFHM line ($\Delta=1$) in the phase diagram with $h=2,\,4,\,\cdots ,\,20$, thus highlighting three different regimes. Note that to improve the statistics, the histograms are constructed over \emph{all} the eigenstates, rather than only for the mid-spectrum states. For weak disorder $h=2$ [panel (a)], the distributions narrow with increasing $L$, in agreement with previous observations~\cite{pal_many-body_2010}. Then, as the disorder strength increases, a new pattern emerges in a regime where otherwise standard observables indicate a rather well-converged MBL-type behavior for these system sizes~\cite{sierant_polynomially_2020,colbois_interaction_2024}: the distributions tend to  broaden slightly, but more strikingly, a fat tail (seemingly power-law) appears  for small ${\ell}^{z}$ and then persists for disorder strengths roughly in the range $h\sim 4-10$, see panels (b-e) (this behavior is best seen for the largest $L$). Finally, as the disorder is further increased, this peculiar regime is followed by a more conventional behavior where the distributions become closer to a normal distribution (in the ${\ell}^{z}$ variable, with an exponential-like tail not emphasized here) as $L$ increases, see panels (f-h). 
Note that the observation of this "fat tail" behavior for $P(\ell^z)$ was not reported in Ref.~\cite{pal_many-body_2010}, which instead mentioned log-normal distributions once $h>4$ in the putative MBL phase. 
\begin{figure}[h!]
    \centering
    \includegraphics[width=\columnwidth]{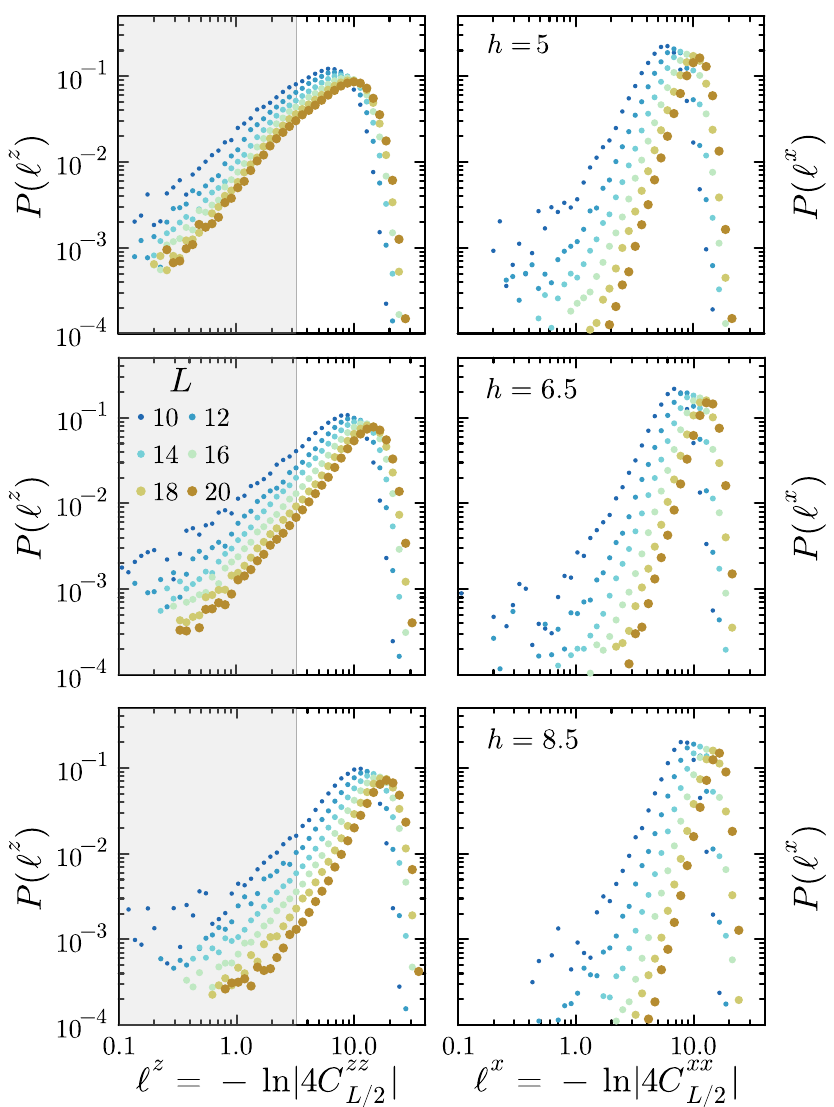}
    \caption{\label{fig:HeavyCorrelations} Distributions  $P(\ell^z)$ (left) and $P(\ell^x)$ (right) for three disorder strengths $h=5,\, 6.5,\,8.5$ along the RFHM line ($\Delta=1$). The histograms, collected from mid-spectrum eigenstates are shown for $L=10,\,12,\,14,\,16,\,18,\,20$. The fat-tail regime for the longitudinal component $\ell^z$ contrasts with the quickly vanishing tail for the transverse one. The region indicated by the shaded gray area corresponds to $\ell^z_* < 3.22$ ($|C^{zz}_{L/2}|>10^{-2}$), and is used to compute the weight $W_*$ in Eq.~\eqref{eq:Wstar}, shown in Fig.~\ref{fig:WCavgDelta1}.}
\end{figure}

{\it Comparison with transverse correlator --- } Fig.~\ref{fig:HeavyCorrelations} compares $P(\ell^x)$ and $P(\ell^z)$ for three representative disorder strengths ($h=5,\,6.5,\,8.5$) in the regime where the fat tail emerges for $\ell^z$ (here, only eigenstates close to the middle of the spectrum are kept). $P(\ell^x)$ is clearly suppressed much faster with system size and does not exhibit a tail for low values of $\ell^x$. In this regime, $P(\ell^x)$ is qualitatively similar to the Gaussian-like behavior observed for $P(\ell^z)$ at stronger disorders (compare to  Fig.~\ref{fig:distrib_overview} (f-h)). The distribution of $\ell^{x}$ translates into a behavior much more similar to the many-body Anderson insulator (Sec.~\ref{sec:AL}), that is, a standard localized regime in which the statistical analysis is not dominated by rare events.

The different shapes of $P(\ell^\alpha)$, and their evolution with increasing system size, provide a qualitative explanation for the differences between average and typical observed in Sec.~\ref{sec:avg_typ}, a point which we discuss more quantitatively below.

\subsection{Quantitative analysis of the fat-tail regime}
\label{sec:fattail}
We quantify the importance of the tail in the distribution of $|C_{L/2}^{zz}|$ (gray shaded area in Fig.~\ref{fig:HeavyCorrelations}) by the weight 
\be
W_{*}=\int_{0}^{\ell^z_*}P(\ell^z){\rm d}\ell^z,
\label{eq:Wstar}
\ee
where $\ell^z_*$ is a size-independent ${\cal{O}}(1)$ cutoff. 
For a log-normal distribution, assuming that the standard deviation (in $\ell^z$) grows slowly enough~\footnote{It is sufficient that $\sigma$ grows slowly enough compared to $\overline{\ell^{z}}$. Here we have $\sigma\sim\sqrt{L}$, see Appendix~\ref{sec:variance}}, $W_{*}$ would vanish exponentially with $L$, such that average and typical correlations would be similar. The situation is quite different in the presence of a fat tail, which leads to a slower decay of both $W_{*}(L)$ and the average correlation function, as we discuss below for two representative cuts.

\subsubsection{$\Delta=1$: horizontal scan}
\label{sec:fattailhorizontal}
\begin{figure}[t!]
    \centering
    \includegraphics[width=\columnwidth]{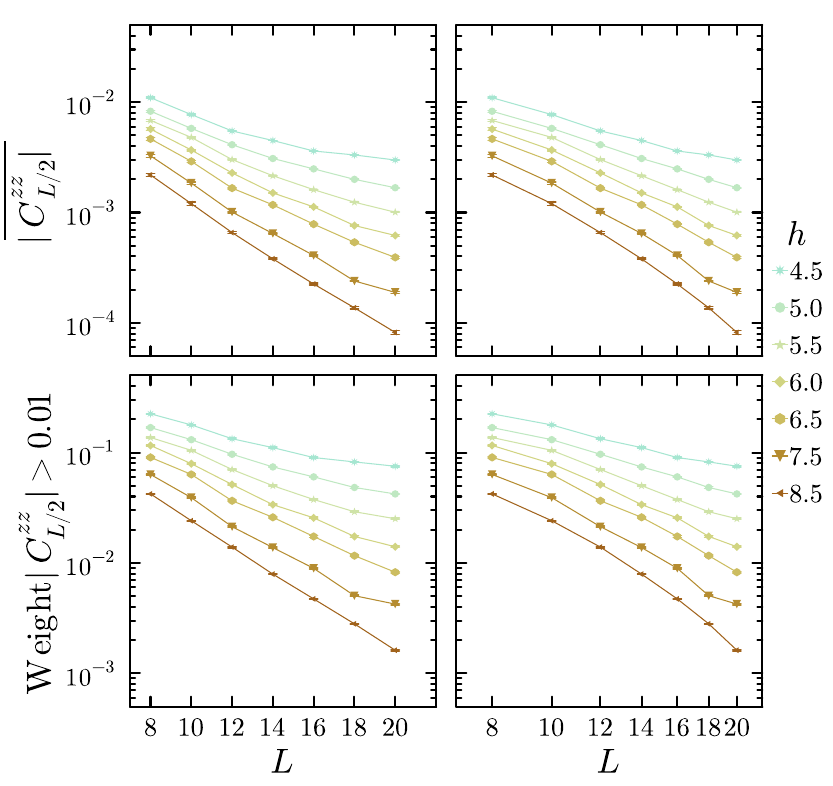}
    \caption{Average midchain longitudinal correlations ${\overline{|C^{zz}_{L/2}|}}$ (top) and large correlation weight $W_*$, Eq.~\eqref{eq:Wstar} (bottom), both plotted against $L$ for $\Delta=1$ and various disorder strengths $h$, as indicated on the plot. The semi-log scale (left) indicates that only the data at $h=8.5$ can be reasonably described by an exponential decay, while the log-log scale (right) shows a plausible power-law decay for most of the other data, see Fig.~\ref{fig:WCavgexponentsDelta1} for a quantitative analysis.\label{fig:WCavgDelta1}}
    \end{figure}

Focusing on the regime $4.5\le h\le 8.5$, we plot in Fig.~\ref{fig:WCavgDelta1} both the weight $W_*$ Eq.~\eqref{eq:Wstar}, computed for $|C_{L/2}^{zz}|> 0.01$ (corresponding to $\ell^z< 3.22$~\footnote{The precise value of $\ell_*^z$ is not crucial, as long as it remains  ${\cal{O}}(1)$ and not too small so that we can numerically resolve the weight $W_*$.}, shaded area in Fig.~\ref{fig:HeavyCorrelations}), and the average correlation, formally defined as
\be
{\overline{|C_{L/2}^{zz}|}}=\frac{1}{4}\int_{0}^{\infty}P(\ell^z)\exp(-\ell^z){\rm d}\ell^z.
\ee
These two quantities show qualitatively similar trends, clearly demonstrating that the average is controlled by the left tail of $P(\ell^z)$, i.e. the weight of large correlations.
For most values of $h$ in this regime, we observe a sub-exponential decay with $L$, presumably better described by a power law, as shown in the log-log plots of Fig.~\ref{fig:WCavgDelta1} (right panels). On the other hand, the data for $h=8.5$ seem to more reasonably obey a pure exponential decay with $L$, as shown by the semi-log plots in Fig.~\ref{fig:WCavgDelta1} (left panels).

To make this analysis more systematic across the entire horizontal cut ($\Delta=1,\,h\le 10$), we fit the decay of both $W_*(L)$ and ${\overline{|C_{L/2}^{zz}|}}$ by exponential forms
\be
W_{*}(L)\sim \exp\left(-\frac{L}{2\lambda_{\rm w}}\right)\, \, {\rm{and}}\,\, {\overline{|C_{L/2}^{zz}|}\sim \exp\left(-\frac{L}{2\xi^{z}_{\rm avg}}\right)},
\label{eq:exp}
\ee
as well as using algebraic laws
\be
W_{*}(L)\sim L^{-\eta_{\rm w}}\quad {\rm{and}}\quad {\overline{|C_{L/2}^{zz}|}}\sim L^{-\eta^{z}_{\rm avg}}.
\label{eq:alg}
\ee
%
As previously done in Sec.~\ref{sec:corrl}, the length scales $\lambda_{\rm w}$, $\xi^{z}_{\rm avg}$, and the power-law exponents $\eta_{\rm w,\,avg}$ are estimated from fits performed over 3 different sliding windows ($L\in [10,\,16] ; [12,\,18]; [14,\,20]$) so that one can estimate the (in)stability of the presumed forms Eqs.~(\ref{eq:exp}, \ref{eq:alg}).

\begin{figure}
    \centering
    \includegraphics[width=\columnwidth]{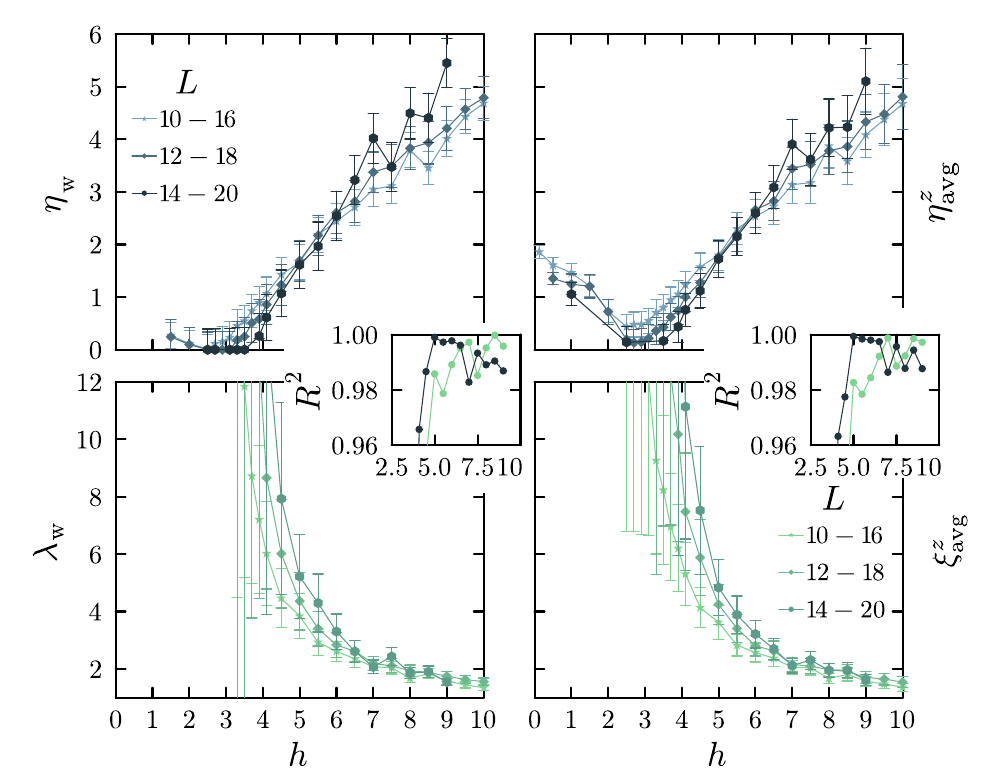}
    \caption{$\Delta = 1$. Characteristic lengths $\lambda_{\rm w}$ and $\xi_{\rm avg}^{z}$ (bottom) and power-law exponents $\eta_{\rm w,avg}$ (top), obtained by fitting the longitudinal  tail weight $W_*$ and correlator ${\overline{|C_{L/2}^{zz}|}}$ to respectively exponential and power-law forms Eq.~\eqref{eq:exp}, and Eq.~\eqref{eq:alg}). These estimates, obtained over different sliding windows ($L\in [10,\,16] ; [12,\,18]; [14,\,20]$), are  plotted against the disorder strength $h$, highlighting three qualitatively distinct regimes (see text). The insets quantify  the qualities of the fitting forms, defined in Eq.~\eqref{eq:Rsquared}.}
     \label{fig:WCavgexponentsDelta1}
\end{figure}

The results are shown in Fig.~\ref{fig:WCavgexponentsDelta1} where we observe a strong similarity between $W_*(L)$ and ${\overline{|C_{L/2}^{zz}|}}$, such that $\lambda_{\rm w}\approx \xi^{z}_{\rm avg}$ and $\eta_{\rm w}\approx \eta^{z}_{\rm avg}$. In addition, the study of fit stability with the sliding windows allows one to distinguish three qualitatively different regimes: {\it (i) }For strong disorder (above $h\sim 8$) the data are best described by an exponential decay, stable with $L$ and with rather short characteristic length scales; {\it (ii) }in the other limit for $h\lesssim 5$, the exponential fits are clearly not stable with $L$, while the power-law fits show a slow $L$-dependent crossover for the exponent $\eta$~possibly at weak disorder towards the value $\eta=1$ expected for ETH (see Sec.~\ref{sec:ETH} and Fig.~\ref{fig:Ergodic1}, this is more clearly seen for $\eta^z_{\rm avg}$); {\it (iii) } finally, the intermediate region $5\lesssim h \lesssim 8$ is not correctly described by an exponential decay, and is best characterized by a rather stable algebraic decay, with exponents $2< \eta< 4$. To confirm this visual estimate of stability of the fits, we compute their coefficient of determination obtained by comparing the residuals of the data $y$ with respect to the fit $f$ from those with respect to the average : 
\begin{equation}
    \label{eq:Rsquared}
    R^2 := 1 - \frac{\sum_i {(y_i -f_i)^2}}{\sum_{i}(y_i -\overline{y})^2}.
\end{equation}
$R^2 = 1$ ($R^2 \ll 1$) indicates a perfect (bad) fit. Despite the noise inherent to the original data, the plateau close to $1$ for $R^2$ for the power-law fits displayed in the insets in Fig.~\ref{fig:WCavgexponentsDelta1} highlights the existence of an intermediate regime where the power-law fits indeed perform better.

\subsubsection{$h=6$: vertical scan}
\label{sec:fattailvertical}
\begin{figure}[t!]
    \centering
    \includegraphics[width=\columnwidth]{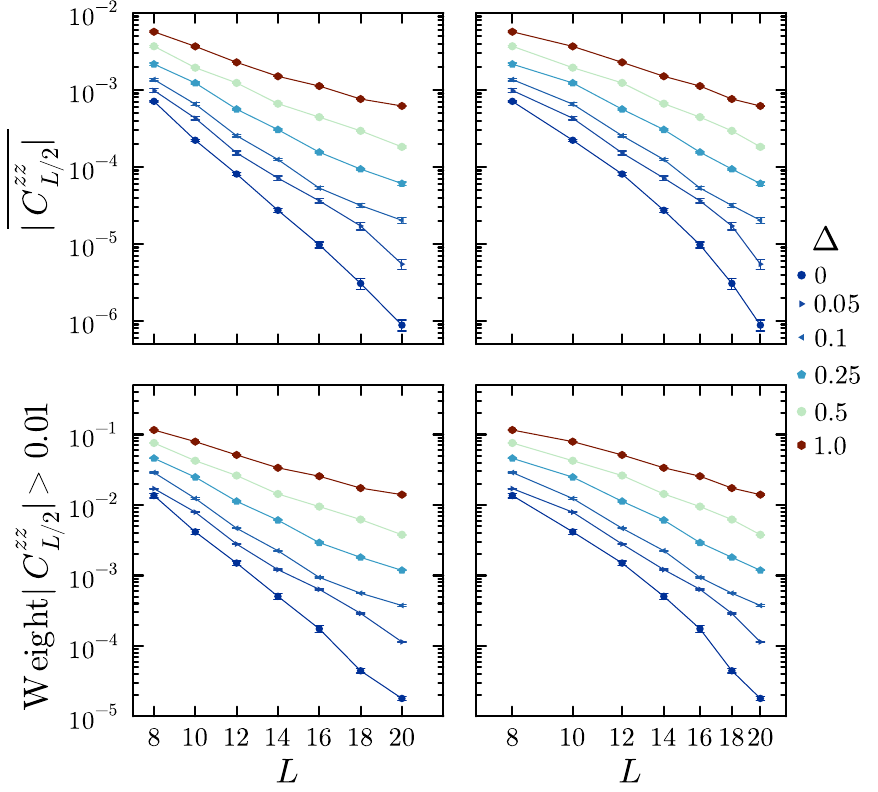}
    \caption{\label{fig:WCavgh6} Same as Fig.~\ref{fig:WCavgDelta1} but for a vertical scan at $h=6$ plotted for $\Delta\in [0,\,1]$.  The semi-log scale (left) indicates that the data for $\Delta < 0.1$ can be described by an exponential decay, while with the log-log scale (right) we observe a reasonnable power-law decay for $\Delta\gtrsim 0.1$. Fig.~\ref{fig:WCavgexponentsh6} shows a quantitative analysis of the fitting parameters.}
\end{figure}

\begin{figure}[t!]
    \centering
    \includegraphics[width=\columnwidth]{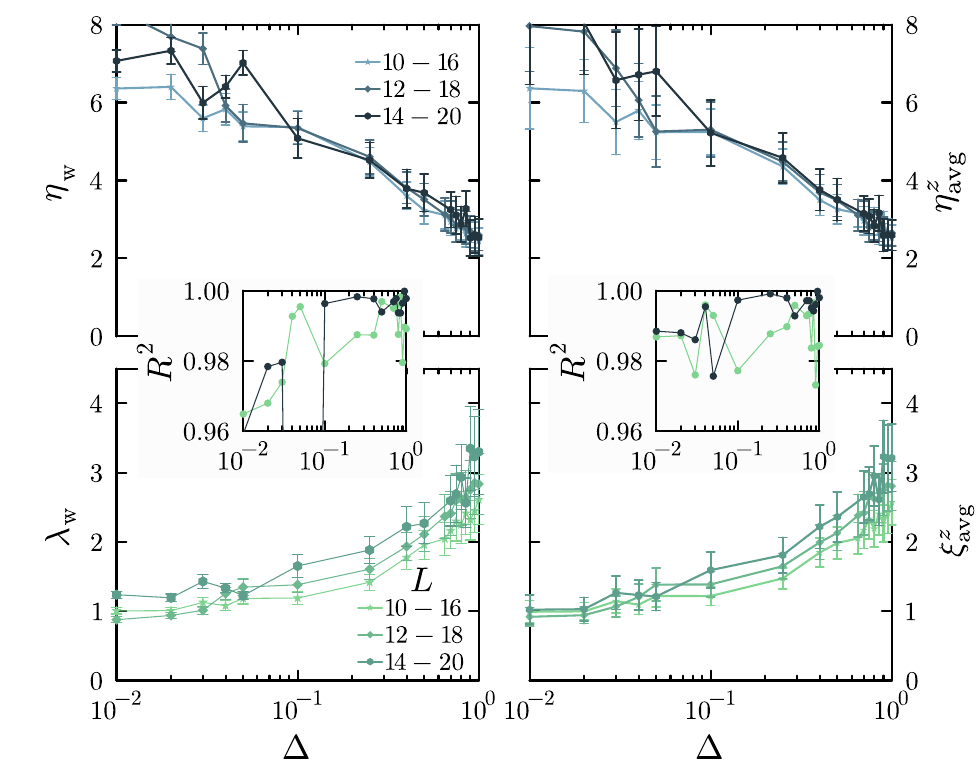}
    \caption{Same as Fig.~\ref{fig:WCavgexponentsDelta1} but for a vertical scan at $h=6$ with $\Delta\in [0,\,1]$. The power-law regime appears to be stable above $\Delta\sim 0.1$, while below that an exponential decay seems to describe the available data slightly better. \label{fig:WCavgexponentsh6}}
\end{figure}

An identical analysis is performed for a vertical scan, examining how $P(\ell^z)$ evolves for a fixed disorder $h = 6$ while $\Delta$ is varied. Fig.~\ref{fig:WCavgh6} shows how the weight $W_*(L)$ and the average correlations ${\overline{|C^{zz}_{L/2}|}}$ decay with $L$ in the range $\Delta\in [0,\,1]$. The results of a quantitative  analysis of the decays, fitted with exponential Eq.~\eqref{eq:exp} and algebraic Eq.~\eqref{eq:alg} forms, are shown in Fig.~\ref{fig:WCavgexponentsh6}, similar to the analysis in Fig.~\ref{fig:WCavgexponentsDelta1} for the RFHM.


    
 Although the error bars are quite large, especially for the power-law exponents $\eta_{\rm w,avg}$ (top panels), we can still conclude that at sufficiently low interaction, roughly below $\Delta \sim 0.1$, we do not observe signs of fat tails, with both $W^*(L)$ and ${\overline{|C^{zz}_{L/2}|}}$ compatible with exponential decay, at least in the range of system sizes accessible to our simulations. This regime would then be adiabatically connected to the many-body Anderson insulator. However, as the interaction strength increases, we observe a qualitative change towards power-law decay at and above $\Delta \sim 0.1$, signaling the onset of the intermediate fat-tail regime. The results in Appendix~\ref{sec:Vertical} support this picture even further.

\subsection{Discussion}
\label{sec:DiscussionLargeCorrelations}
With the above analysis, we have clearly identified a new regime where the midchain longitudinal correlations follow a fat-tailed distribution for $\ell^z\to 0$, presumably power-law. Along the RFHM line ($\Delta=1$), this roughly occurs for $5\lesssim h \lesssim 8$, i.e. in a regime where otherwise standard eigenstate observables indicate MBL physics ~\cite{sierant_polynomially_2020,colbois_interaction_2024}. An important consequence is that in this region, typical and average correlations have qualitatively different decays:
\bea
{\overline{|C^{zz}_{L/2}|}} &\propto& L^{-\eta_{\rm avg}}\\
\exp\left({\overline{\ln|C^{zz}_{L/2}|}}\right)&\propto& \exp\left({-\frac{L}{2\xi_{z}^{\rm typ}}}\right),
\eea
suggesting that certain types of rare events with large correlation dominate the mean, leading to a behavior clearly distinct from that of the typical value. The other scan for which we have also presented a detailed analysis at $h=6$ similarly displays such an intermediate window, readily observed above $\Delta\sim 0.1$ (for lower values of $\Delta< 0.1$, this regime is not observed within our system sizes). Importantly, we did not find trace of a similar regime for the transverse correlator $C^{xx}_{L/2}$, which instead displays more conventional behaviors (Gaussian-like in the ergodic phase, log-normal like at larger disorder).

The fact that typical and average longitudinal correlations show such different trends suggests the absence of self-averaging for this particular observable. 
Such unconventional behavior reminds us of the case of Infinite Randomness Fixed Point (IRFP) criticality~\cite{fisher_random_1992,fisher_random_1994,fisher_critical_1995}, for which we briefly recall the peculiar properties of the correlation functions at distance $r$. They are broadly distributed, with a fat tail describing rare but large correlation events associated with long-range singlets (aka the random singlet phase). As a striking consequence, the {\it{average}} decays algebraically $C_{\rm avg}(r)\sim r^{-\eta}$, since it is dominated by ${\cal{O}}(1)$ correlations occurring with a probability $\sim r^{-\eta}$, see Refs.~\cite{young_numerical_1996,laflorencie_crossover_2004} for numerical evidence. At the same time, the {\it{typical}} correlation, controlled by a scale-invariant part of the distribution~\cite{fisher_distributions_1998}, decays much faster, like a stretched exponential $C_{\rm typ}(r)\sim \exp(-A\sqrt{L})$. Although IRFP physics is usually associated with a strong disorder type renormalization group flow~\cite{igloi_strong_2005}, it is not clear that such a framework provides a good description of MBL physics~\cite{pekker_hilbert-glass_2014,slagle_disordered_2016,you_entanglement_2016}. For example, Refs.~\cite{sahay_emergent_2021,moudgalya_perturbative_2020,laflorencie_topological_2022} concluded for ${\mathbb{Z}}_2$-symmetric models that at high energies, the non-interacting IRFP is destabilized by any weak interactions towards a delocalized ergodic phase, contrasting with the zero-temperature properties where instead IRFP remains stable against finite interactions in the ground state~\cite{chepiga_resilient_2024}. For the present U(1)-symmetric XXZ model, the non-interacting limit is the many-body Anderson insulator, which is not described by IRFP physics, but rather by a standard finite-disorder fixed point, which has been argued~\cite{mard_strong-disorder_2014} to be analogous to the off-critical (Griffiths) regime of the random transverse-field Ising chain~\cite{fisher_critical_1995,young_numerical_1996}. Therefore, we do not expect finite interaction to emulate any kind of IRFP physics, but our numerical results strongly suggest that rare events may still play a key role.

Remarkably, qualitative differences between average and typical decay have also been reported in apparently quite different situations. In the context of classical spin glasses, Ref.~\cite{morone_large_2014} discussed the relevance of large deviations of correlation functions in random magnets using the example of the random-field Ising model on the Bethe lattice, where they found rare events with anomalously strong correlation functions driven by a few pairs of strongly correlated spins.
In a series of advances on the Anderson transition problem in random graphs, Refs.~\cite{garcia-mata_scaling_2017,garcia-mata_two_2020,garcia-mata_critical_2022} proposed that rare events dominate the physics in the localized regime. In particular, they numerically found qualitatively different behavior for the decays of single-particle two-point correlation functions over the random graph, a behavior that may be related to our observations, although the correspondence between the many-body problem and random graphs is known to be highly non-trivial~\cite{tarzia_many-body_2020,roy_fock-space_2020, roy_localizatino_2020,pietracaprina_hilbert-space_2021,roy_fock_2024, Herre_2023, Tikhonov_2021, Scoquart_2024} and deserves more careful investigation.

\section{Summary, discussions and outlooks}
\label{sec:SummaryOutlook}
We summarize here our main results and connect them with the relevant literature. The basis for our comprehensive study of correlators was laid in section~\ref{sec:warmup} with the limiting cases, where their behavior is well understood. The ergodic regime at low disorder shows different natures for the midchain correlators: algebraic decay with system size $L$ for the longitudinal $C_{L/2}^{zz}$, due to the magnetization preserving nature of ${\cal{H}}_{\Delta}$ in Eq.~\eqref{eq:H}, versus the exponentially decaying transverse component $C_{L/2}^{xx}$. In the non-interacting many-body Anderson case, both correlators decay exponentially with associated typical length scales that follow the localization length and respect a hierarchy $\xi^x_{\rm typ} > \xi^z_{\rm typ}$.

Section~\ref{sec:Overview} contrasts the exponential decay of transverse correlators observed throughout the phase diagram with the richer longitudinal behavior. While for sufficiently strong disorder $C^{zz}_{L/2}$ also decays exponentially, upon decreasing disorder we find a quite unconventional intermediate regime with a broad and asymmetric distribution of systemwide correlation $P(\ln|C^{zz}_{L/2}|)$ that appears well before the ergodic phase.  In this interesting regime, we report very pronounced differences between average and typical midchain correlators. In particular, the average is no longer well described by a stable exponentially decay, while the typical is (note that the latter will eventually also show signs of instability as the disorder is further reduced, as previously reported in Ref.~\cite{colbois_interaction_2024}). 

By focusing in Section~\ref{sec:LargeCorrelations} on large correlation events using the logarithmic variable $\ell^z=-\ln(4|C^{zz}_{L/2}|)$, we can explore the large deviation statistics in this intermediate disorder regime, where we reveal the existence of a fat-tailed probability distribution (presumably a power law) for strong correlators $|C^{zz}_{L/2}|$ (small values of $\ell^z$), an effect that is clearly visible in Fig.~\ref{fig:distrib_overview}. These rare but strong correlations are at the origin of the non-trivial qualitative difference observed between typical and average decays, as also reported in Fig.~\ref{fig:xis}, where the effective correlation lengths, $\xi_{\rm typ}^z(L)$ and $\xi_{\rm avg}^z(L)$, show very contrasting behaviors. This intermediate "fat-tail" regime, in which systemwide correlations are dominated by rare events, exhibits a rather large extension in the disorder-interaction plane (green region in the phase diagram, Fig.~\ref{fig:phasediag}), typically nearly as large as the ergodic region. Beyond this regime, for larger disorder, we report more conventional exponentially suppressed long-range correlations, consistent with a more standard real-space localization.

The existence of a region of the phase diagram, or a regime in the parameter range, presumably inside the MBL phase, with unusual (not-strictly localized) properties has been advocated in some previous works, albeit not always at the same location or not delimited by the same markers~\footnote{The fat-, power-law, tails that we report for the probability $P(\ell^z)$ should not be confused with the ``heavy-tails'' of Ref.~\cite{colmenarez_statistics_2019} which finds power-law behavior for the RFHM for $P(|C_r^{\alpha \alpha}|)$ for intermediate values of $|C|$, at \emph{strong} disorder (deep in the MBL regime), and for small distances $r\leq 4$ and \emph{both} directions $\alpha=x,z$}.  An earlier exact diagonalization analysis~\cite{lim_many-body_2016} argued on the existence of a critical regime for $h\in [3,4]$ based on the power-law decay of the average correlator $C^{zz}_r$,  which we instead attribute to the crossover to the ergodic regime in this disorder range, as already visible for the {\it typical} correlator represented in Fig.~\ref{fig:overview}~(a). Ref.~\cite{herviou_multiscale_2019} considered the distribution of sizes of {\it entanglement} clusters across the phase diagram and found that the probability of small 2-sites entanglement clusters remains finite for disorder values where we observe the fat-tails, but finds that the two sites are at relatively short distances. Similarly, Ref.~\cite{hemery_identifying_2022} obtains relatively local {\it correlation} clusters. This contrasts with the long-range nature of the enhanced correlations that we find. Interestingly, Ref.~\cite{villalonga_characterizing_2020}  analyzes the probability distribution of (the log of) the quantum mutual information (QMI) between two distant spins (closely related to $C^{zz}$ in case of magnetization conservation) for the RFHM. One of the main results is a stretched exponential behavior (as a function of distance $r$) for the typical QMI of open chains in an intermediate disorder regime, whereas our typical data at maximal distance $r=L/2$ are best fit with an exponential decay (albeit in some cases with a growing correlation length).  Ref.~\cite{villalonga_characterizing_2020} also studies events associated to large QMI, but to the best of our understanding does not report power-law tails in the range where we find them, but rather a maximum around $h \simeq 3$ (identified as the transition). We note nevertheless the positive skewness in the distribution of the log of the QMI, in the range where we see fat tails. 
Anomalous dynamical behaviors (in dynamics after a quench, or in transport) have also been reported~\cite{gopalakrishnan_low_2015,barisic_dynamical_2016,prelovsek_density_2017,weiner_slow_2019,chanda_time_2020,nandy_dephasing_2021,sierant_challenges_2022,evers_internal_2023} 
in a similar disorder regime where we find fat-tails for strong correlators, and it would be interesting to understand if these two observations could be related.

Our finding of the enhancement of strong long-distance correlations could be interpreted as a direct witness of long-range many-body resonances, extensively argued~\cite{gopalakrishnan_low_2015, garratt_local_2021,crowley_constructive_2022,garratt_resonant_2022,morningstar_avalanches_2022,Ha_many-body_2023} to be the final precursor to the ergodic phase, and as such complements other (direct or indirect) supports for the identification of long-range many-body resonances in microscopic MBL models~\cite{gopalakrishnan_low_2015,geraedts_many-body_2016,khemani_critical_2017, kjall_many-body_2018, colmenarez_statistics_2019,villalonga_eigenstates_2020, villalonga_characterizing_2020, garratt_local_2021,garratt_resonant_2022,ghosh_resonance_2022,morningstar_avalanches_2022,long_phenomenology_2023,Ha_many-body_2023}.

Considering the logarithmic variable $\ell_z$ effectively zooms on anomalous strong long-distance correlators. 
With a similar objective but a different approach, Ref.~\cite{biroli_large-deviation_2024} proposes a way to put a biased emphasis on long-distance many-body resonances and also find an intermediate regime (between the ergodic and deeply localized) where such rare long-distance resonances dominate the physical behavior (albeit in a different microscopic model). In a related spirit, Ref.~\cite{morningstar_avalanches_2022} considers the extreme statistics of the QMI (the maximal QMI taken over {\it all} eigenstates at the largest distance for open chains) as a marker of the appearance of at least one systemwide resonance, identified at a disorder strength $h_{\rm swr} \simeq 8$  for the RFHM. This value is apparently close to the one where, for $\Delta=1$, we begin to observe power laws for $P(\ell^z)$, but we note that taking the maximum QMI over the entire many-body spectrum should provide an upper bound on the extent of this regime. Our results suggest, at least for periodic systems free of edge effects, that this upper bound should be larger.

Our study does not shed light on the exact anatomy~\cite{roy_fock-space_2020,roy_fock_2024} of the resonant states that give rise to these strong distant correlators, even though we note that 
that simultaneous strong $C^{zz}_{L/2}$ and vanishing $C^{xx}_{L/2}$ suggests the involvement of (at least two) spin flips in the bulk due to the total magnetization conservation in our samples. Further work is needed to identify the form and range distribution of these strong correlation events, as well as to make connections with phenomenological approaches to many-body resonances \cite{garratt_local_2021,crowley_constructive_2022,garratt_resonant_2022,crowley_mean-field_2022, long_phenomenology_2023}, avalanche models~\cite{de_roeck_stability_2017,thiery_many-body_2018,crowley_avalanche_2020,crowley_constructive_2022,crowley_mean-field_2022,morningstar_avalanches_2022,Ha_many-body_2023}, and other renormalization group / effective approaches to the transition~\cite{potter_universal_2015, thiery_many-body_2018,thiery_microscopically_2017,dumitrescu_kosterlitz-thouless_2019, morningstar_renormalization-group_2019,morningstar_many-body_2020,garcia-mata_two_2020,roy_spectral_2024}.

Although our results are effectively limited to finite systems, one can nevertheless try to elaborate possible scenarios for the intermediate regime in larger chains, and in particular how it relates to its surrounding ergodic and MBL phases. 

1)~The fat tail of large systemwide correlations may be an early sign of a regime that eventually becomes ergodic at larger scales, a scenario that falls into the phenomenology of the prethermal MBL~\cite{long_phenomenology_2023}.
However, it remains to be understood why in this regime many-body spectroscopy adheres so strongly to Poisson statistics, as also raised
by Biroli {\it{et al.}}~\cite{biroli_large-deviation_2024}. 

2)~ As an alternative scenario, one could imagine that the observed power-law decay of the average systemwide correlations is only a transient finite-size effect before an asymptotic exponential decay signaling a stable conventional MBL phase at larger $L$. Although we cannot rule it out, this scenario seems unlikely from our data, which instead show that the boundary of this regime drifts toward larger disorder with increasing $L$.

3)~Finally, as a third possibility, one may expect the power-law decay of the average correlations to persist at larger sizes, giving rise to an intermediate stable phase where ergodicity is broken but real-space localization is not total. 

This clearly paves the way for interesting future research aimed at exploring these different scenarios. In that respect, tensor network approaches targeting excited states~\cite{khemani_obtaining_2016, lim_many-body_2016, yu_finding_2017, villalonga_exploring_2018} may help clarify the fate of the average systemwide correlations for larger systems. Their bias towards less entangled states could, however, be a significant limitation~\cite{hemery_identifying_2022, yu_finding_2017}.

As a final remark, we believe that further studies of long-distance correlators in microscopic models (with or without magnetization conservation, including in dynamical setups) would provide very useful insights, since they are natural, experimentally accessible~\cite{leonard_probing_2023}, and interpretable probes of many-body resonances. This real-space perspective on localization could form a basis for deeper understanding, and reveal the key role of extreme statistics in the stability of MBL.

\acknowledgments
We thank G. Biroli, W. Chen, G. Lemari\'e and M. Tarzia  for very fruitful discussions. This work has been partly supported by  the EUR grant NanoX No. ANR-17-EURE0009 in the framework of the ”Programme des Investissements d’Avenir”, is part of HQI initiative (www.hqi.fr) and is supported by France 2030 under the French National Research Agency award number ANR- 22-PNCQ-0002, and also benefited from the support of the Fondation Simone et Cino Del Duca. JC acknowledges support from the Singapore Ministry of Education Academic Research Fund Tier 2 (grant MOE-T2EP50222-0005). We acknowledge the use of HPC resources from CALMIP (grants 2023-P0677) and GENCI (project A0150500225), as well as of the PETSc~\cite{petsc-user-ref,petsc-efficient}, SLEPc~\cite{slepc-toms,slepc-users-manual}, MUMPS~\cite{MUMPS1,MUMPS2} and Strumpack~\cite{Strumpack} sparse linear algebra libraries.

\vskip 1cm
\setcounter{section}{0}
\setcounter{secnumdepth}{3}
\setcounter{figure}{0}
\setcounter{equation}{0}
\renewcommand\thesection{S\arabic{section}}
\renewcommand\thefigure{S\arabic{figure}}
\renewcommand\theequation{S\arabic{equation}}

\appendix

\begin{center}
    \bfseries\Large Appendix
\end{center}

In these appendices, we provide details for several points of the main text. Appendix~\ref{sec:fermions} deals with the free-fermion calculations.  Appendix~\ref{sec:RS} gives more details on correlations in random vectors modeling ETH. Appendix~\ref{sec:variance} provides a brief discussion of the variance of the correlation distributions in the interacting case. Finally, Appendices~\ref{sec:Horizonal} and~\ref{sec:Vertical} show the results of a similar analysis as in the main text, but for other cuts in the phase diagram.  Appendix~\ref{sec:Horizonal} gives the  results for $\Delta = 0.25, 0.5$ and~$0.75$ while Appendix~\ref{sec:Vertical} provides further results at constant fields $h = 4$~and~$6$. We discuss the evaluation of the boundary of the intermediate regime (green dots in Fig.~\ref{fig:phasediag}), and of the associated error bars.
\section{Many-body Anderson insulator}
\label{sec:fermions}

\subsection{Transformation to free fermions}
The XXZ Hamiltonian Eq.~\eqref{eq:H} can be re-written in terms of spinless fermions in a random potential through the Jordan-Wigner transformation~\cite{jordan_uber_1993}:
\begin{equation}
n_i = S_{i}^{z} +1/2, \quad  c_i = \exp \left( i\pi \sum_{j=1}^{i-1} n_j \right) S_i^{-},  
\end{equation}
with the term in the exponent corresponding to the Jordan-Wigner string. In one dimension, this string cancels out for nearest-neighbor hopping, yielding
\begin{equation}
    H_f = \sum_i \left[\frac{1}{2} \left( c_i^\dagger c_{i+1}{\vphantom{\dagger}} + c_{i+1}^\dagger c_i{\vphantom{\dagger}} + 2\Delta n_i n_{i+1} \right) + h_i n_i\right] + H_{\cal B} +C
\end{equation}
where the boundary term $H_{\cal B}$ comes  from the Jordan-Wigner string, and $C$ is a constant energy shift. 
In the Anderson case, $\Delta = 0$ and the fermionic Hamiltonian is quadratic and can be exactly diagonalized by new fermionic operators of the form
\begin{equation}
    b_m = \sum_{i=1}^{L}\phi_{m}(i) c_i.
\end{equation}
Thus we have $H_f = \sum_{m = 1}^L \mathcal{E}_{m} b_m^{\dagger} b_m{\vphantom{\dagger}}$ with $\mathcal{E}_m$ the \emph{single-particle} energy associated with the $m^{\rm  th}$ exponentially localized orbital $\phi_m$. The conservation of the magnetization corresponds to a conserved number of spinless fermions, with the $S^{z}_{\rm tot} = 0$ sector of the spin chain corresponding to $L/2$ occupied orbitals $\{m_{\mu}\}_{\mu = 1}^{L/2}$. To form the many-body wavefunction, we randomly select the occupied orbitals in such a way as to target many-body states at energy density above the ground state $\epsilon = (E-E_{\min})/(E-E_{\max}) = 0.5 $~\cite{hopjan_detecting_2021}. For a given disorder realization, the many-body energy is given by $E = \sum_{j = 1}^{L/2} \mathcal{E}_{m_{j}}$, with $E_{\min}$ ($E_{\max}$) corresponding to the ground (top) state.

\subsection{Spin-spin correlations}

Both the longitudinal and transverse correlations can be expressed in terms of fermions and then evaluated using Wick's theorem, as detailed in particular in Refs.~\cite{henelius_numerical_1998, lieb_two_1961}. For $i = j$, we have $\langle S_i^{\alpha} S_{i}^{\alpha}\rangle = \frac{1}{4}$. For $i \neq j$ the longitudinal correlations, 
\begin{align}
    \langle S_i^{z} S_j ^{z} \rangle &= \langle (c_i^{\dagger} c_i^{\vphantom{\dagger}} -1/2)(c_j^{\dagger} c_j^{\vphantom{\dagger}} -1/2) \rangle\\
    &= \langle c_i^{\dagger} c_i^{\vphantom{\dagger}} c_j^{\dagger} c_j^{\vphantom{\dagger}} \rangle  - \frac{1}{2} \langle n_i + n_j\rangle + 1/4  
\end{align}
so that the connected $zz$ correlations read
\begin{equation}
    C^{zz}_{ij} = \langle c_i^{\dagger} c_i^{\vphantom{\dagger}} c_j^{\dagger} c_j^{\vphantom{\dagger}} \rangle - \langle n_i\rangle \langle n_j\rangle = -\langle c_i^{\dagger}c_j^{\vphantom{\dagger}} \rangle \langle c_j^{\dagger} c_i^{\vphantom{\dagger}} \rangle,
\end{equation}
where in the last equality we used Wick's decoupling. This is readily expressed in terms of the \emph{occupied} orbitals as in Eq.~\eqref{eq:Czz1} and results in purely negative $C^{zz}$~\cite{colmenarez_statistics_2019}.

Ref.~\cite{henelius_numerical_1998} shows how to evaluate the transverse correlations. We need to define $A_i = c_i^{\dagger} + c_i^{\vphantom{\dagger}}$, $B_i = c_i^{\dagger} - c_i^{\vphantom{\dagger}}$. Both these operators intervene when using Wick's theorem, with the only non-zero contraction given by:
\be
    \langle A_i B_j \rangle = -\langle  B_j A_i\rangle  = \delta_{ij} - 2 \left(\sum_{\mu = 1}^{L/2} \phi_{m_{\mu}}^{\ast}(j) \phi_{m_{\mu}}(i) \right).
\ee
Defining $G_{ij} := \langle B_i A_j \rangle$, we can obtain the transverse correlations from the determinant: 
\be
 \langle S_i^{+} S_j^{-} \rangle = \frac{1}{2} \begin{vmatrix}
G_{i,i+1} & G_{i,i+2} & \cdots & G_{ij} \\
G_{i+1,i+1} & G_{i+1,i+2} & \cdots & G_{i+1,j} \\
\vdots & \vdots & \ddots & \vdots \\
G_{j-1,i+1} & G_{j-1,i+2} & \cdots & G_{j-1,j}
\end{vmatrix}
\ee
which is easy to evaluate numerically.

\subsection{Link between longitudinal and transverse correlations}

In Sec.~\ref{sec:AL}, we argued that $\xi^{x}_{\rm{typ}} \approx 2\xi^{z}_{\rm{typ}}$ but also noticed that $\xi^{x}_{\rm avg} < 2 \xi^{z}_{\rm avg}$.  Here, we use the simplified picture of simple log-normal distributions to  explain this observation.

Fig.~\ref{fig:AL2} shows the joint distributions of the logarithm ${P} (\ln|C^{xx}_{L/2}|, \ln|C^{zz}_{L/2}|)$.
The maximal probability lies on the line corresponding to  $\ln|C^{zz}_{L/2}| = 2\ln|C^{xx}_{L/2}| +2\ln 2$, implying a factor two between the longitudinal and transverse typical correlation lengths. Yet, the distribution is quite broad, compatible with a standard deviation of the form (see also Appendix~\ref{sec:variance})
\begin{equation}
    \sigma (\ln|C^{\alpha \alpha}_{L/2}|) \sim  \beta_{\alpha}\sqrt{ (L + o(\ln L))}.
    \label{eq:sigma}
\end{equation}

To see how this can impact the average behavior, we first rescale the correlation functions using the Pauli matrices, such that 
${\C}^{\alpha\alpha}=4C^{\alpha\alpha}$.
Then, using $\ell^{\alpha} = - \ln|{\C}^{\alpha\alpha}_{L/2}|$, we can write $\C^{\alpha\alpha}_{\text{typ}} =\exp(\overline{-\ell^{\alpha}})$ and $\C^{\alpha\alpha}_{\text{avg}} = \overline{\exp\left(-\ell^{\alpha}\right)}$ (the subscripts $L/2$ have been dropped to simplify the notations).
For sufficiently strong $h$, we assume that each sample/eigenstate expectation yields
\be
\ell^{z} = 2\ell^{x},
\ee
obviously giving the expected result for the typical correlations
\be
\C^{zz}_{\text{typ}} = \exp(\overline{-2\ell^{x}}) = [\exp(\overline{-\ell^{x}})]^2 = (\C^{xx}_{\text{typ}})^2.
\ee
However, for the average we have
\begin{align}
\C^{zz}_{\text{avg}} &= \overline{\exp(-2\ell^{x})} = \overline{[\exp(-\ell^{x})]^2}\\
&= (\C^{xx}_{\rm{avg}})^2 + \sigma^{2}_{|\C^{xx}|}\neq (\C^{xx}_{\rm{avg}})^2 
\end{align}
where we have used the fact that ${\overline{\left(\C^{xx}\right)^2}}=(\C^{xx}_{\rm{avg}})^2 + \sigma^{2}_{|\C^{xx}|}$.

\begin{figure}[h!]
	\centering
	\includegraphics[width=\columnwidth]{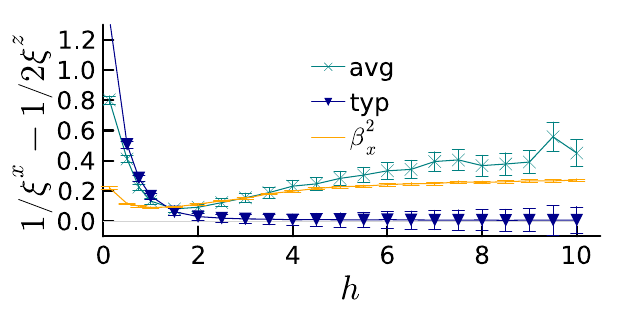}
 \caption{\label{fig:avgcorrlengths} Relation between the longitudinal and transverse correlation lengths at $\Delta = 0$. While the typical lengths are related by a factor 2 when the disorder is sufficiently strong, there is a further disorder-dependent correction for the average, qualitatively captured by Eq.~\eqref{eq:avgcorrlengthAL} despite the approximations.  }
\end{figure}

If we \emph{assume} a log-normal $P(\C^{\alpha\alpha})$ (implying that $\ell^{\alpha}$ follows a normal distribution), we obtain
\be
\sigma^2_{\ell^\alpha}=\ln\left(1+\frac{\sigma^2_{\C^{\alpha\alpha}}}{(\C_{\avg}^{\alpha\alpha})^2}\right).
\label{eq:sigCsigl}
\ee
Ignoring the subleading log corrections, we use
\be
\sigma^2_{\ell^\alpha}= \beta_{\alpha}^2 L,
\ee
(in the large-$L$ limit) and invert Eq.~\eqref{eq:sigCsigl}, to  obtain
\be
\C^{zz}_{\text{avg}}=(\C^{xx}_{\rm{avg}})^2 \exp\left(\beta_{x}^2 L\right).
\ee
Finally, this gives the following relation between the inverse average correlation lengths from midchain correlations
\be
 \frac{1}{\xi^{x}_{\rm avg}} \approx \frac{1}{2 \xi^{z}_{\rm avg}} + \beta_{x}^2.
 \label{eq:avgcorrlengthAL}
\ee
Thus the factor 2 present for the typical correlation lengths is replaced by a slowly varying term for the average correlation lengths, as shown in Fig.~\ref{fig:avgcorrlengths}. Recall that Eq.~\eqref{eq:avgcorrlengthAL} is valid for log-normal distributions, which do not perfectly describe the transverse and longitudinal correlations in AL. Despite this limitation, the agreement in Fig.~\ref{fig:avgcorrlengths} is somewhat reasonable.

\section{Random state}
\label{sec:RS}
\begin{figure}[b!]
	\centering
\includegraphics[width=\columnwidth]{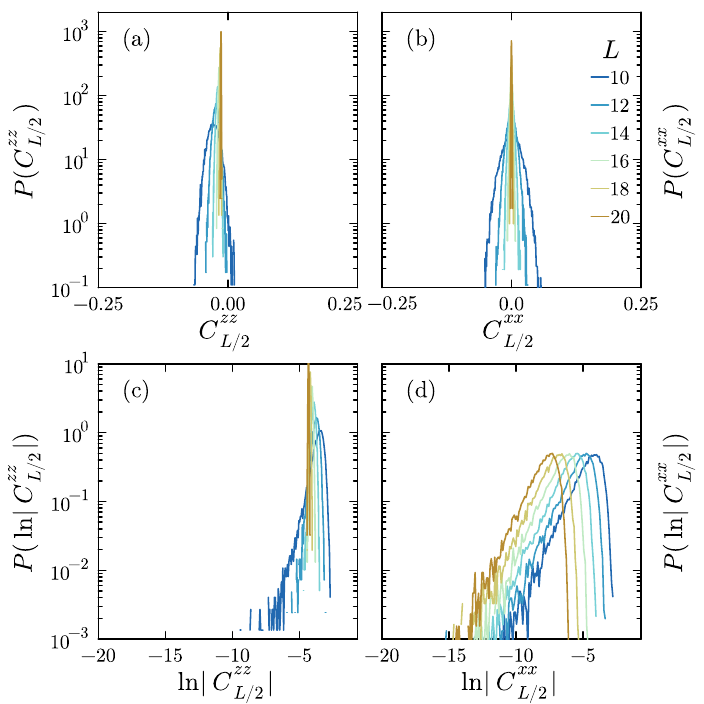}
\caption{\label{fig:RandomVector} Distributions of the midchain spin-spin correlations for a random vector in the zero magnetization sector, to compare to Fig.~\ref{fig:Ergodic2}.}
\end{figure}
At finite interaction, the limit of weak disorder, i.e. deep in the ergodic regime, was discussed in Sec.~\ref{sec:ETH}. In Fig.~\ref{fig:Ergodic2}, we presented the distributions of the correlations at $\Delta = 1, h = 1$. For reference, in Fig.~\ref{fig:RandomVector} we show the distributions of the longitudinal and transverse correlations and of their logarithms for random vectors sampled from the $S^{z}_{\rm tot} = 0$ sector. Panels (a) and (b) clearly show Gaussian distributions with a reducing variance. Additionally, it is easy to see that longitudinal correlations are shifted to negative values due to the sum rule and that they are almost all negative. 
Panels (c) and (d) of Fig.~\ref{fig:RandomVector} illustrate the strong contrast created by this simple sum rule for the distributions of the logarithms. The transverse correlations have self-similar distributions which only get shifted with increasing system sizes with a constant variance. In contrast, the longitudinal correlations get more and more peaked around the typical value, which decays as a power-law of the system size (see Fig.~\ref{fig:Ergodic1}).

Thus, the random vector captures very well the most important effects present at $\Delta = 1, h = 1$ (Fig.~\ref{fig:Ergodic2}). The only slight difference is that in Fig.~\ref{fig:Ergodic2}, $P(C^{zz}_{L/2})$ has more weight on positive correlations than in the random vector case. This results in the tail at small correlations for $P(\ln|C^{zz}_{L/2}|)$, which however reduces with increasing system sizes.

\section{Variance}
\label{sec:variance}

\begin{figure}[h!]
	\centering
	\includegraphics[width=\columnwidth]{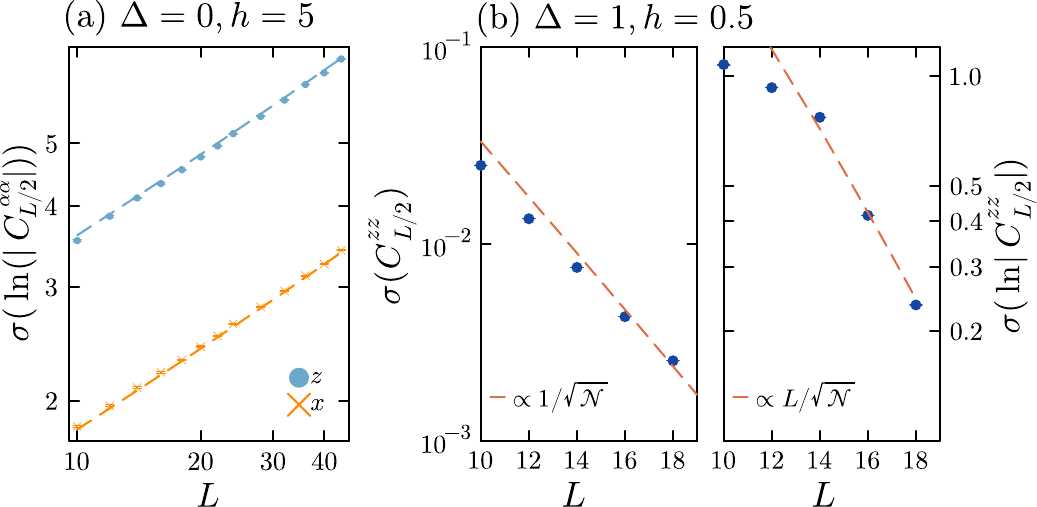}
	\caption{\label{fig:VarianceBoth}(a) Standard deviation of the distributions of the log longitudinal and transverse correlations at $\Delta = 0$, and fits to square-root growth with logarithmic corrections. (b) Decay of the variance of the longitudinal correlations, deep in the ETH regime ($h = 0.5, \Delta = 1$).}
\end{figure}

We now mention results for the standard deviations of the logarithm of the longitudinal and transverse correlations.
The limiting cases are shown in Fig.~\ref{fig:VarianceBoth}. In the non-interacting AL insulator, fitting the standard deviation growth with simply a power-law yields $\sigma \sim L^{-0.43}$ at $h = 5$. However, this behavior can be ascribed to additive logarithmic corrections to the variance~\cite{fisher_distributions_1998}, as stated in Eq.~\eqref{eq:sigma}. Fig.~\ref{fig:VarianceBoth}(a) shows that the numerical results are compatible with such corrections. At $h = 5$, we find: 
\begin{equation}
      \sigma (\ln|C^{\alpha \alpha}_{L/2}|) =   \beta_{\alpha}\sqrt{ (L + 2\ln L)} \quad \beta_{z} \approx 2\beta_{x} \approx 0.943 
\end{equation}
At weak disorder and strong interaction in the ergodic regime, $\sigma(\ln|C^{xx}_{L/2}|)$ is constant as expected from ETH (see the distributions in Fig.~\ref{fig:Ergodic2}), whereas $\sigma(\ln|C^{zz}_{L/2}|)$ reaches a maximum and decays with increasing system sizes. More precisely, Fig.~\ref{fig:VarianceBoth}(b) shows that the decay is roughly related to the square-root of the Hilbert space dimension, see also Section~\ref{sec:ETH}). 

\begin{figure}[t!]
	\centering
	\includegraphics[width=\columnwidth]{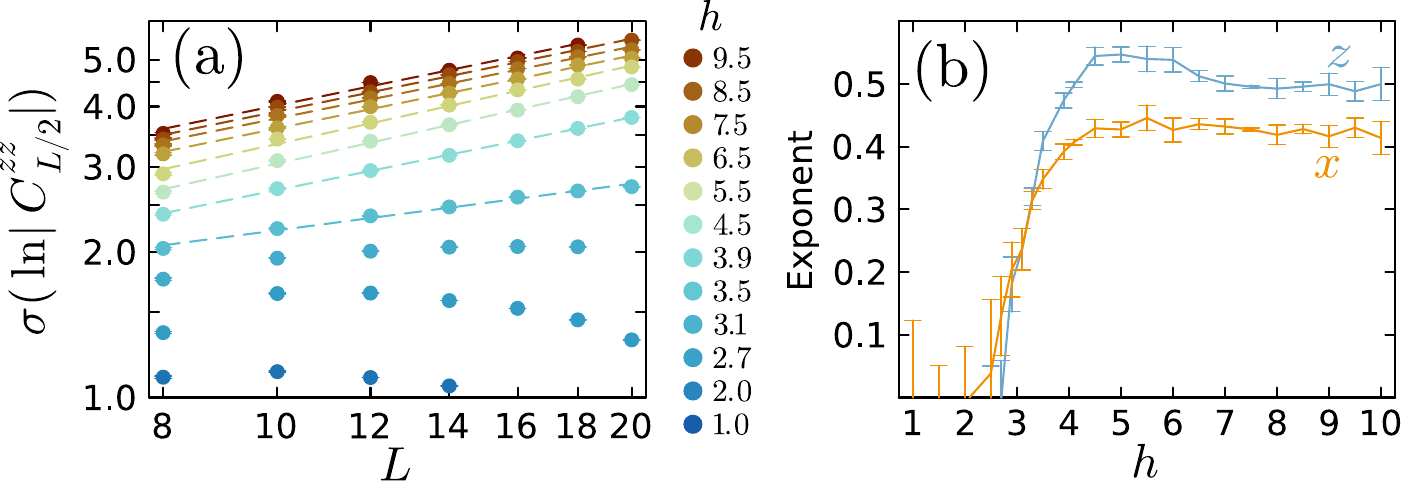}
	\caption{\label{fig:VarDelta1} (a) Standard deviation of the distributions of the log longitudinal correlations at $\Delta = 1$. For disorders $h \gtrsim 3.9$,  the standard deviation grows as a power-law with system size $L$. (b) Exponent of the  power-law fit for even sizes $L = 8$ to $L = 20$ for both components. At strong disorder for the longitudinal correlations, it is compatible with the expected square-root increase. The transverse standard deviation can be accounted for a logarithmic corrections to a square-root growth. }
\end{figure}

Fig.~\ref{fig:VarDelta1}(a) presents the standard deviation of the logarithm of the longitudinal correlations at $\Delta = 1$. 
For sufficiently strong disorder, both $\sigma(\ln|C^{zz}_{L/2}|)$ and $\sigma(\ln|C^{xx}_{L/2}|)$ (not shown) grow as a power-law of the system size. The effective exponents in Fig.~\ref{fig:VarDelta1}(b) show a very clear square-root behavior for the longitudinal correlations, and a sightly lower power for the transverse correlations, compatible with the logarithmic corrections present also in the non-interacting results. 

Taken together with the evolution of the typical correlations, these results are compatible with those of Ref.~\cite{pal_many-body_2010} for the random-field Heisenberg chain, where their $\sigma$ stands for $\sigma(\ln|C^{zz}|) /\overline{\ln|C^{zz}|}$. However, our findings in the main text highlight the essential role played by the fat-tail at strong longitudinal correlations, an effect that is not captured by the scaling of $\sigma(\ln|C^{zz}|)$.

\section{Constant interactions}
\label{sec:Horizonal}

\begin{figure}
	\centering
\includegraphics[width=0.65\columnwidth]{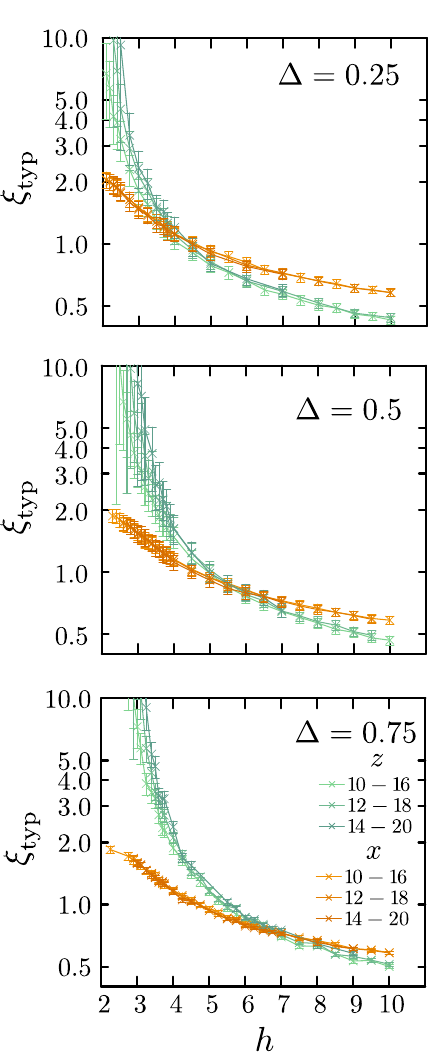}
	\caption{\label{fig:corrlengthoverview}Overview of the midchain correlation lengths throughout the phase diagram (compare to Fig.~\ref{fig:xis}). The lengths associated with the midchain average correlations are shown in Fig.~\ref{fig:exponentsoverview}. The middle panel was already shown in~\cite{colbois_interaction_2024}.}
\end{figure}

\begin{figure}
	\centering
\includegraphics[width=0.97\columnwidth]{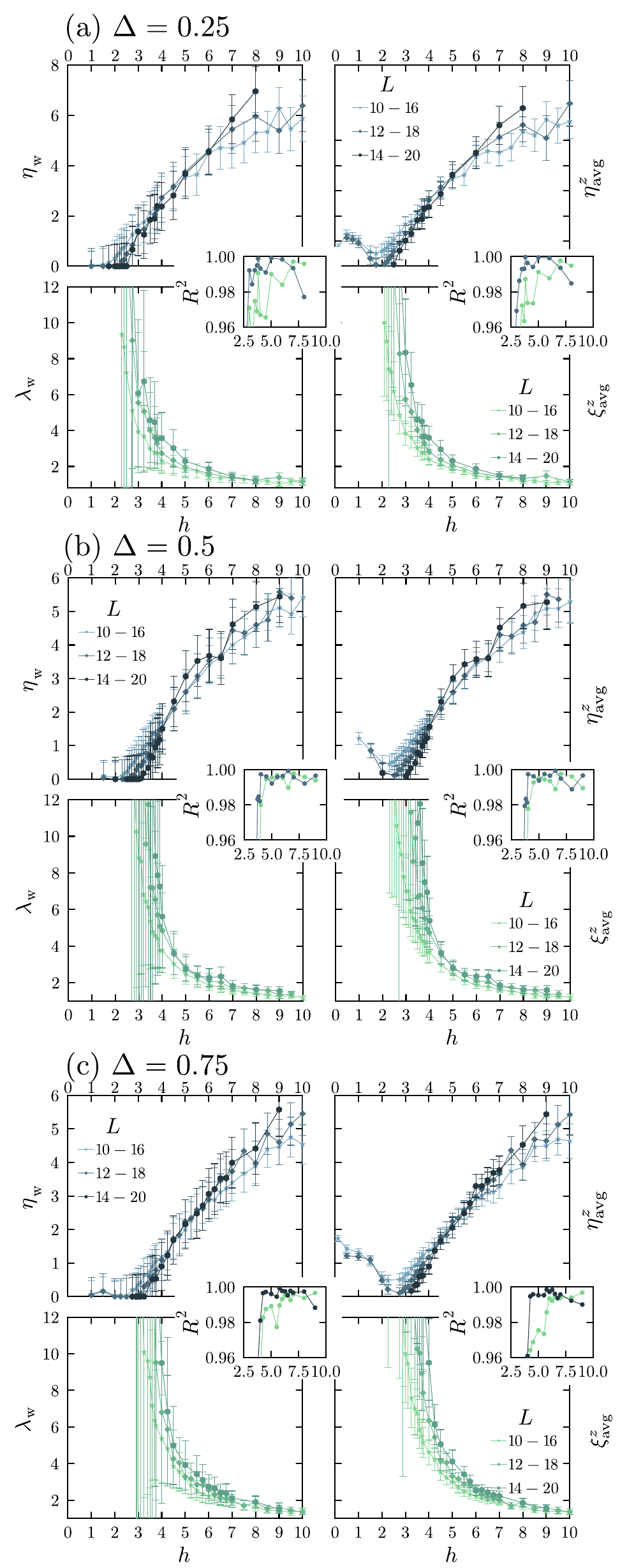}
	\caption{\label{fig:exponentsoverview} Overview of the midchain correlations throughout the phase diagram (compare to Fig.~\ref{fig:WCavgexponentsDelta1}). Left column: decay of the weight of large correlations fitted to a power-law (top panels) and exponential (bottom panels). Right column : power-law (top) and exponential (bottom) fits to the decay of the average $|C^{zz}_{L/2}|$. }
\end{figure}

\begin{figure*}
	\centering
	\includegraphics[width=0.9\textwidth]{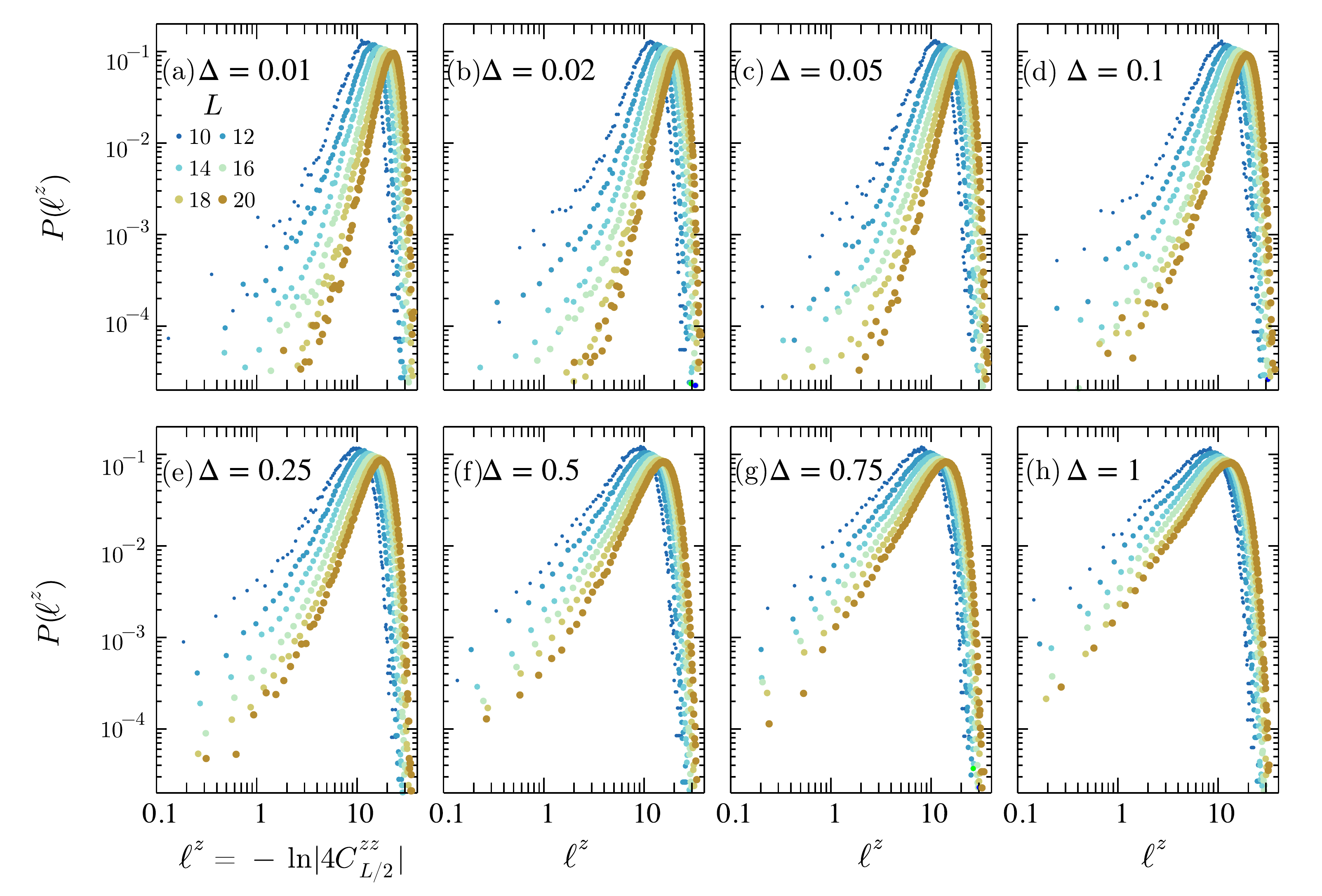}
	\caption{\label{fig:h6histograms} Evolution of the distributions $P(\ell^{z})$ along a constant field line $h=6$ for interactions from $\Delta = 0.01$ to $\Delta = 1$. As in Fig.~\ref{fig:HeavyCorrelations}, and unlike Fig.~\ref{fig:distrib_overview}, the histograms are collected across about 60 eigenstates near $\epsilon = 0.5$ (and less than 1\% of the spectrum, for small sizes) and thousands of disordered samples. Small values of $\ell^z$ correspond to very large correlations (close to the maximal possible value), while large values of $\ell^z$ correspond to extremely small correlations.}
\end{figure*}

In the main text, we have focused on the correlation functions in the random-field Heisenberg model. Here, we provide an analysis of other constant interaction lines, at $\Delta = 0.25, 0.5$~and~$0.75$. We focus on the typical and average decays as well as on the weight of large correlations (fat-tails).

Fig.~\ref{fig:corrlengthoverview} provides an overview of the evolution of the \emph{typical} longitudinal and transverse correlation lengths. It can be compared to Figs.~1 and~3 of Ref.~\cite{colbois_interaction_2024} as well as Fig.~\ref{fig:xis} of the present work. Altogether, these figures support the following important results (see also main text):

{\it (i)} The typical longitudinal length grows quickly with interactions, contrasting with the almost constant transverse correlation length.

{\it (ii)} Throughout the presumably MBL phase (as given by standard observables), this results in a change of ordering from $\xi^z_{\rm typ} < \xi^x_{\rm typ}$ at strong disorder and weak interactions (similar to the AL limit) to $\xi^z_{\rm typ} > \xi^x_{\rm typ} $ at  weaker disorder and stronger interactions (Fig.~\ref{fig:corrlengthoverview}). 

{\it (iii)} An instability of the longitudinal typical length shows up in roughly the region where  $\xi^z_{\rm typ} \sim \xi^x_{\rm typ} $.

We suggested in the main text that this instability could be a weak signature of rare, large correlations, that are much more manifest in the average correlations. Here, we provide additional evidence supporting this scenario throughout the phase diagram. Despite the large errors, Fig.~\ref{fig:exponentsoverview} signals a region of power-law decay of the average correlations, supporting further our results in Sec.~\ref{sec:LargeCorrelations}. 

Using these results to trace out the intermediate regime in Fig.~\ref{fig:phasediag} is not easy for the following reason. At very strong disorder ($h \gtrsim 9$), the average correlations fluctuate a lot. Importantly, this purely numerical effect shows up already in the Anderson localized case for strong disorder $h \gtrsim 8$ -- in some cases, the decay can \emph{seem} algebraic even though a pure exponential is expected (not shown here). The limit on small correlations imposed by machine precision together with the low probability of large correlations imply that sampling effects become much more important. Upon introducing interactions, the challenging region starts from stronger disorders (as the correlations grow). Hence, this difficulty does not undermine our results at disorders below $h\sim 8-9$ in the phase diagram.

From Fig.~\ref{fig:phasediag}, we extract the boundary and corresponding errors as follows:
\begin{enumerate}
    \item For $\Delta = 0.25$ in Fig.~\ref{fig:exponentsoverview}(a), the fits to power-law decays of both the average correlations and of the weight of large correlations are consistently good until $h \leq 6$ (presumably even $h \leq 6.5$), while $h \geq 7$ is on the edge with an exponential fit seeming slightly better. The error on the right is estimated in light of the fact that the decay at $h = 8$ is clearly exponential, suggesting that the change of behavior occurs before. 
    \item For $\Delta = 0.5$ (panel (b)), the situation is much more delicate, with a lot of fluctuations in the longitudinal correlations. However, the $R^2$ (Eq.~\eqref{eq:Rsquared}) suggests that the correlations decay is slightly better described by a power-law at least until $h \lesssim 6.5$, which provides the lower bound on the error in Fig.~\ref{fig:phasediag}, while most points beyond this are compatible with an exponential decay. The very large error in the phase diagram corresponds to the unexpected, probably accidental, agreement with a power-law decay at $h = 9, \Delta = 0.5$. 
    \item For $\Delta = 0.75$ in panel (c), the situation is again much clearer, with a power-law decay until $h \lesssim 7$, an intermediate situation for $h \sim 8$ followed by an exponential decay at $h \sim 9$: we place the upper limit for the end of the power-law regime before this value. 
\end{enumerate}

In our view,  despite the challenges, these results present a consistent picture for the phase diagram. 
The next section supports this even further by providing a similar analysis, this time for fixed disorder strength and variable interaction.

\section{Constant fields}
\label{sec:Vertical}
\subsection{Fat-tail regime}
In Sec.~\ref{sec:fattailvertical}, we have presented a quantitative analysis of the decays of the average correlations and the weights of the fat-tails. Here, we provide some further  evidence, qualitative for $h = 6$, and quantitative  for $h = 4$.

First, Fig.~\ref{fig:h6histograms} gives a qualitative overview of the distributions at constant fields. The distributions for weak interactions [panels (a)-(c)] exhibit a strikingly different behavior from strong interactions [panels (f)-(h)]. In the first panels, the large correlations are suppressed very quickly with increasing system sizes, in agreement with the exponential decay of the associated weights Figs.~\ref{fig:WCavgh6} and~\ref{fig:WCavgexponentsh6}.  Instead, at strong interactions, a very clear power-law tail develops at small $\ell^z$, i.e. large correlations, again agreeing with the power-law decay of the weights of large correlations shown in Fig.~\ref{fig:WCavgh6}. 
In the intermediate values of interactions, $0.1 \lesssim \Delta \leq 0.25$, there are some signatures of the presence of a fat-tail regime. While it is difficult to identify the precise limiting value of the interactions from only a qualitative inspection of the distributions, the integrated weight in Fig.~\ref{fig:WCavgexponentsh6} indicates clearly that all this region also corresponds to a power-law decay. Taken together, the results at $h = 6$ do provide ample evidence for the existence of a wide power-law regime for the decay of the average correlations. 

\begin{figure}[h!]
    \centering
    \includegraphics[width=\columnwidth]{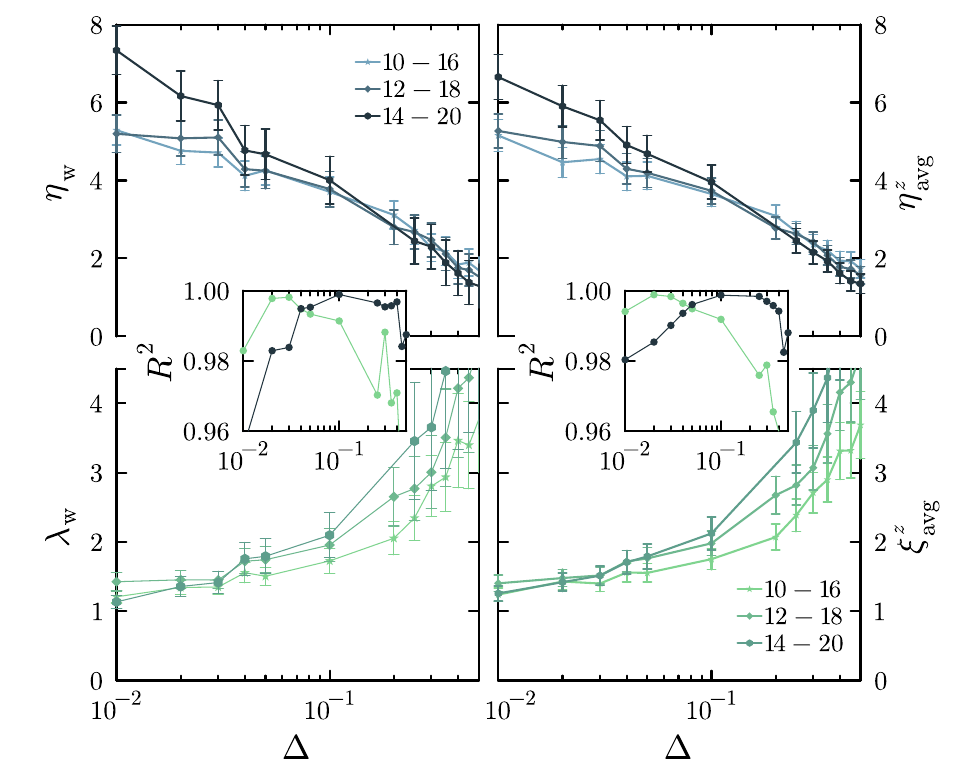}
    \caption{Same as Fig.~\ref{fig:WCavgexponentsh6} but for $h = 4$. For this field, we note that for the finite sizes presented here, all standard observables show non-ergodic behavior for any $\Delta \leq 0.5$. Note that in Fig.~\ref{fig:phasediag}, the extrapolated critical disorder out of the ergodic phase for $\Delta = 0.25$ is of order $h_c = 3.7 \pm 0.6$. This suggests that above $\Delta \sim 0.25$, the power law decay of the average correlations may be due to the presence of the ergodic phase.}
    \label{fig:WCavgexponentsh4}
\end{figure}

To further establish that a truly \emph{exponential} decay of average longitudinal correlations can be found at weaker (but non-zero) interactions, we complement our results at $h = 6$ with evidence at $h = 4$, shown in Figure~\ref{fig:WCavgexponentsh4}. For $\Delta \lesssim 0.05$, this plot reveals a consistent exponential decay of the average correlations, with the top panels clearly ruling out a power-law decay.  In contrast, for $\Delta \geq 0.1$, the results are consistent with a power-law. 

\subsection{Weak interaction limit}
\label{sec:departure}
\begin{figure}[h!]
    \centering
    \includegraphics[width=\columnwidth]{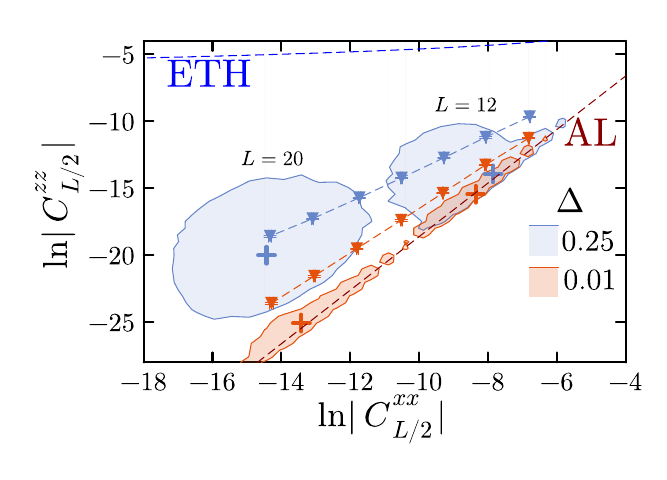}
    \caption{Contours at half-maximum weight of $\mathcal{P}(\ln|C^{xx}_{L/2},\ln|C^{zz}_{L/2}|)$ at $h = 6$, for sizes and interaction strengths as indicated in the legend, with the location of maximal probability indicated by a "+" sign. The down triangles illustrate the typical correlations for even sizes $L = 8, \dots, 20$. \label{fig:h6cut} }
\end{figure}

We finally point out an intriguing effect of the weak interactions on  the joint distributions of mid-chain correlation, ${P}(\ln|C^{xx}_{L/2}|,\ln|C^{zz}_{L/2}|)$. In Sec.~\ref{sec:AL} and in particular Fig.~\ref{fig:AL2}, we have seen that the majority of the weight accumulates along the ``AL line'' given by Eq.~\eqref{eq:XXZZAL}. In Sec.~\ref{sec:Overview} and Fig.~\ref{fig:overview}, we have seen very clearly that even at strong interactions and strong disorder (in the presumably MBL regime), the correlations instead have much broader distributions, noticeable in the contours at half-maximum. The typical correlations also follow a very different trend as the one shown by the AL line. In Fig.~\ref{fig:h6cut}, we put the emphasis on how this behavior arises at weak but non-zero interaction. 

For very weak interactions (orange, $\Delta = 0.01$), the contour at half maximum for $L= 12$ accumulates on the AL line. However, the relation between transverse and longitudinal correlations (orange triangles) exhibits a slightly different slope. Upon increasing the system size, eventually, the contour also starts detaching from the AL accumulation line. This shows the extreme sensitivity to vanishingly small but finite $\Delta$ on the Anderson insulator.

At $\Delta = 0.25$ (blue), the slope of the typical correlations is markedly different, and the contours at half-maximum cover a much broader reagion. They clearly detach from the AL accumulation line for increasing system sizes. For $L = 12$, the most probable point ("+" symbol) is still on the AL line, but similarly to what happens in Fig.~\ref{fig:overview} at $\Delta = 1, h = 8.5$, it jumps away for a large enough size. 

\newpage\bibliography{mbl_correlations_long_paper,notes}

\end{document}